\documentclass[preprint,prd, notitlepage, nofootinbib]{revtex4-2}
\usepackage{amsmath,amssymb}
\usepackage{graphicx}
\usepackage{grffile}
\usepackage{color}
\usepackage{animate}
\usepackage{mathtools} 
\usepackage{bm}
\usepackage{multirow}

\usepackage{ulem} 

\begin{document}

\preprint{OU-HET-1181}
\title{
Why magnetic monopole becomes dyon in topological insulators
}
\author{Shoto Aoki, Hidenori Fukaya, Naoto Kan, Mikito Koshino, and  Yoshiyuki Matsuki}
\affiliation{Department of Physics, Osaka University, Toyonaka, Osaka 560-0043, Japan}
\begin{abstract}
  The Witten effect predicts that a magnetic monopole 
acquires a fractional electric charge inside topological insulators.
  In this work, we give a microscopic description of this phenomenon, 
as well as an analogous two-dimensional system with a vortex.
We solve the Dirac equation of electron field both analytically 
in a continuum space and numerically on a lattice,
by adding the Wilson term and smearing the gauge field within a finite range 
to regularize the short-distance behavior of the system.
Our results reveal that the Wilson term induces
a strong positive mass shift,
creating a domain-wall around the monopole/vortex.
This small, yet finite-sized domain-wall localizes the chiral zero modes 
and ensures their stability through the Atiyah-Singer index theorem,
whose cobordism invariance is crucial in explaining 
why the electric charge is fractional.
\end{abstract}
\maketitle
\newpage
\section{Introduction}
\label{sec:intro}

The magnetic monopole has been a fascinating object in particle physics.
It explains the quantized nature of electric charges of elementary particles~\cite{Dirac:1931kp}, 
describes duality of the Maxwell theory under exchange of electric and magnetic fields 
and appears as a stable solitonic state in grand unified theories~\cite{tHooft:1974kcl,Polyakov:1974ek}.
Although it has not been discovered yet,
even its absence plays a role
in supporting the inflation scenario of the early universe,
where the density of the monopoles became extremely small~\cite{Guth:1980zm}.

In this article our focus is on one of remarkable theoretical predictions about a monopole
that it gains a nonzero electric charge in the so-called $\theta$ vacuum with a nonzero angle.
This prediction arises from a direct coupling between electric and magnetic fields
and is known as the Witten effect \cite{Witten:1979ey}.
The effect has provided 
significant insights into understandings of quantum field theory and string theory
\cite{Callan:1982ah,Callan:1982au,Rubakov:1982fp,Harvey:1996ur,Alvarez-Gaume:1997dpg, Pretko:2017xar,Yamamoto:2020phl, Abe:2022nfq}.

Nontrivial $\theta$ dependence can appear from
a non-zero expectation value of the axion field background,
which has been actively discussed in various studies~\cite{Sikivie:1984yz, Hidaka:2020iaz,Fukuda:2020imw,Agrawal:2022yvu, Hamada:2022bzn,Choi:2022fgx}.
However, our universe is essentially in the $\theta=0$ vacuum, 
where the $CP$ symmetry
with $C$ transformation (exchanging electrons and positrons)
and $P$ transformation (exchanging left-handed and right-handed spins)
is manifestly maintained\footnote{In this work we ignore a tiny $CP$ violation coming from  the weak interaction.}.
The only known exception is topological insulator \cite{Fu:2006djh,PhysRevB.76.045302,Moore:2006pjk,Qi:2008ew,Schnyder:2008tya,PhysRevB.79.195322}, 
which effectively induces the vacuum angle $\theta=\pi$, 
and the $CP$ and time-reversal ($T$) symmetries are protected in a nontrivial way.

Then an interdisciplinary question between particle and condensed-matter physics
arises: what will happen 
to a magnetic monopole when it is put inside a topological insulator?
According to the prediction in Ref.~\cite{Witten:1979ey},  
it will become a dyon with a half-integral electric charge.
A macroscopic understanding of the Witten effect in
condensed-matter physics was reported in Refs.~\cite{Qi:2008ew,Sasaki_2014,Zirnstein2020TopologicalME}.
In this work, 
we aim to provide a microscopic 
understanding of the mechanism behind the capture of
the electric charge by the monopole.
While it is difficult to identify the origin of the electric charge 
from  the $\theta$ term added in the Maxwell theory,
it is almost obvious that the electron
is the only candidate inside condensed-matter.
Then we should address the following issues:
1) how a monopole can capture an electron state, 
2) why this phenomenon does not occur in normal insulators,
and 3) why the charge is not an integer but a half-integer.

In this study, we solve the Dirac equation
in the presence of a monopole
both analytically in continuum and numerically on a lattice.
Additionally, we investigate a vortex in a two-dimensional topological insulator,
which also captures a half-integral charge \cite{Hou:2006qc,Mesaros:2012hu,Lee:2019rfb},
and we find that the microscopic mechanism of
the electron capture is the same for both cases.
To distinguish between normal and topological insulators and to eliminate UV divergence near the monopole/vortex, 
we introduce the Wilson term in the Dirac operator. 
We also smear the magnetic charge/flux distribution
in a range of a finite radius $r_1$ to study
the origin of the nontrivial boundary condition \cite{deResende:2017hxw}
at the singularity of the gauge fields.
Our computation is UV finite everywhere,
allowing us to safely take the $r_1\to 0$ limit without any ambiguity.

Our study reveals dynamical formation of a 
domain-wall around the monopole/vortex with a finite radius.
Inside the domain-wall, the electron enters the trivial phase,
due to an additive mass shift caused by the dense magnetic field in the Wilson term.
This means that the neighborhood of the monopole/vortex becomes a normal insulator island.
In contrast to the standard argument where 
the electron zero mode is captured by the zero-dimensional defect of the gauge field
\cite{Kazama:1976fm,Goldhaber:1977xw,Callias:1977cc, Yamagishi:1982wp, Grossman:1983yf, Yamagishi1983Fermion-monopole,  zhao2012magnetic,Tyner:2022qpr}
the bound state 
can be identified as a chiral edge mode \cite{Jackiw1976Solitons,CALLAN1985427Anomalies,KAPLAN1992342AMethod,Shamir1993Chiral,Furman1995Axial}
on the codimension-one domain-wall around the monopole/vortex.
The chiral boundary condition arises as a dynamical consequence of
the domain-wall formation, rather than an artificial boundary condition imposed by hand.

We find that the chiral modes near the monopole/vortex
obey a massless Dirac equation on the domain-wall
where a strong curvature induces
a gravitational connection \cite{PhysRevB.87.205409, Aoki:2022cwgv, Aoki:2022aez}.
This gravitational effect leads to a gap in the Dirac Hamiltonian,
which is counteracted in our system 
by the magnetic field of the monopole/vortex resulting in the zero eigenvalue.
Moreover, we demonstrate that the chiral zero modes 
are topologically linked
via the Atiyah-Singer(AS) index \cite{atiyah1963index,AtiyahSinger1986TheIndex1} 
(or mod-two AS index \cite{AtiyahSinger1971TheIndex5} for the vortex) 
to the total magnetic flux inside the domain-wall.
Thus, the existence of the zero modes is topologically protected.

The cobordism invariance of the AS index requires
the existence of an additional zero mode 
localized at the surface of the topological insulator
with the opposite chirality,
which is also topologically protected. 
This has been empirically observed in previous 
numerical studies \cite{Rosenberg:2010ia, zhao2012magnetic, Tyner:2022qpr, Aoki:2022aez}.
The paired two zero modes must mix while maintaining the $\pm$ symmetry
of the Dirac Hamiltonian spectrum.
Since only one of the two (near) zero modes is occupied
in the half-filling state, and only half of its amplitude is
located near the monopole/vortex, the zero mode mixing
naturally explains the half integral electric charge in 
the vicinity of the monopole/vortex.

Furthermore, the creation of the domain-wall offers an intriguing 
reinterpretation of the conventional $\theta$ vacuum approach in the Maxwell theory.
The defect in the $\theta$ term does not necessarily require a point-like
magnetic charge. 
If we could introduce a thin solenoid with a sufficiently dense magnetic flux
into a topological insulator, we might be able to
verify experimentally that the magnetic monopole-``like'' configuration
captures half of an electric charge.
In this study, we simulate this system on a lattice
by gradually increasing the magnetic flux of the solenoid.
We find that the charge is pumped not from the bulk but 
from the outer surface of the topological insulator,
where the Dirac string becomes a bridge between the two domain-walls.

The rest of the paper is organized as follows.
In Sec.\ref{sec:EFT}, we review the original semi-classical
effective theory approach with the $\theta$ term and
Chern-Simons term in the Maxwell theory in three and two dimensions, respectively.
Our analysis starts from a vortex
in a two-dimensional topological insulator in Sec.~\ref{sec:vortex}
and continues to our main target: a monopole in a three-dimensional
topological insulator in Sec.~\ref{sec:monopole}.
In Sec.~\ref{sec:lattice}, we numerically confirm
pairing of the zero modes: one located at the monopole and
the other at the surface of the topological insulator.
Then we try to reinterpret the original effective theory approach
in Sec.~\ref{sec:EFT2}, simulating a solenoid
whose end produces the essentially same phenomenon as the Witten effect.
In Sec.~\ref{sec:summary}, we give a summary and discussion.

\section{Effective theory description of monopole/vortex gaining electric charges}
\label{sec:EFT}

First we review the standard understanding how the monopole becomes
a dyon when the effective theory contains the $\theta\neq 0$ term. 
Let us consider a Dirac fermion determinant with a mass $m$ in four dimensions
regularized by a Pauli-Villars(PV) field\footnote{
In order to fully regularize the theory, we
need multiple PV fields. However, for the
tree-level computation in this work, one bosonic spinor field is enough. 
}
with a mass $M_{\rm PV}$,
\begin{align}
  \label{eq:ZPV}
Z = \det \left(\frac{D+m}{D+M_\text{PV}}\right),
\end{align}
where $D=\gamma^\mu D_\mu=\gamma^\mu(\partial_\mu + iA_\mu)$ is the Dirac operator
with the $U(1)$ electro-magnetic gauge field $A_\mu$. 
Here we take $M_\text{PV}$ positive and $m<0$ in a $T$-symmetry protected
topological insulator.

When we perform an axial $U(1)$ rotation to flip the mass sign,
the anomaly produces the $\theta=\pi$ term,
\begin{align}
  \label{eq:Ztheta}
Z = \det \left(\frac{D+|m|}{D+M_\text{PV}}\right)e^{iS_\text{top.}},
\end{align}
where the topological action $S_\text{top.}$ is
\begin{align}
S_\text{top.}
 = \pi\frac{1}{32\pi^2}\int d^4x F_{\mu\nu}\tilde{F}^{\mu\nu},
\end{align}
where $F_{\mu\nu}$ is the field strength of the $U(1)$ gauge field,
and $\tilde{F}^{\mu\nu} =\epsilon^{\mu\nu\rho\sigma}F_{\rho\sigma}$ with the 
anti-symmetric Levi-Civita tensor $\epsilon^{\mu\nu\rho\sigma}$.
Reducing the cutoff $M_\text{PV}\to |m|$, or equivalently integrating out the fermions,
the remaining effective action is the $\theta=\pi$ term. 

The $\theta$ term modifies the Maxwell equation in the vacuum to
\begin{align}
\partial_\mu F^{\mu\nu} = - \frac{\theta}{8\pi^2}\partial_\mu \tilde F^{\mu\nu}.
\end{align}
The $\nu=0$ component relates the divergence of the electric field $\bm{E}$
to that of the magnetic field $\bm{B}$.
The Gauss law around the monopole is then 
\begin{align}
  \label{eq:Gauss}
q_e=\int d^3x \nabla \cdot \bm{E} = -\frac{\theta}{4\pi^2}\int d^3x \nabla \cdot \bm{B} = -\frac{\theta q_m}{2\pi}.
\end{align}
When the magnetic charge $q_m$ is nonzero,
the electric charge $q_e$ is also nonzero.
In particular, when the monopole has a unit magnetic charge $q_m=1$,
the electric charge\footnote{From the $2\pi$
  periodicity of $\theta$,  $-1/2$ mod $\mathbb{Z}$ is allowed.} at $\theta=\pi$ is $-1/2$.

In a similar way, the $2+1$-dimensional effective action of a single Dirac fermion in
a topological phase is the Chern-Simons action with level $k=1$,
which modifies the Maxwell equation to
\begin{align}
\partial_\mu F^{\mu\nu} = - \frac{k}{8\pi^2} \epsilon^{\nu\rho\sigma}F_{\rho\sigma},
\end{align}
and the vortex with flux $\alpha$ gains an electric charge through
the Gauss law
\begin{align}
 \label{eq:Gauss2D}
  q_e=\int d^2x \nabla \cdot \bm{E} = -\frac{k}{2\pi}\int d^2x
  \epsilon^{0\rho\sigma}F_{\rho\sigma}= -k\alpha.
\end{align}
When $\alpha=1/2$ and $k$ is odd, the electric charge is fractional.

This effective theory description is quite simple,
but cannot answer the following questions: 1) What is the origin of electric charge?
2) If the origin is the electron field, why is it bounded to monopole/vortex?
and 3) Why is the electric charge fractional?
In the following sections, we try to answer these questions
revealing the microscopic features of the electron field in the presence of
the monopole/vortex.

\section{Two-dimensional vortex}
\label{sec:vortex}

We start our analysis from a two-dimensional 
massive Dirac fermion,
which represents a model of electron fields
in a two-dimensional topological insulator.
As will be shown below, the microscopic mechanism explaining how a vortex 
gains an electric charge is essentially the same
as what happens on a monopole in a three-dimensional topological insulator.
Summary of analytic results and some numerical analysis on a
lattice were already presented in {Refs.}~\cite{Aoki:2022cwgv,Aoki:2022aez} 
by two of the authors.

\subsection{A naive Dirac equation}

First we review the standard analysis of the Dirac equation
in the literature \cite{Khalilov:2014rka}
to clarify the problems and our goals.
Let us consider a two-dimensional fermion system described by
the Dirac Hamiltonian with a mass $m$.
We put a $U(1)$ gauge flux located at the origin,
whose vector potential is given by
\begin{align}
A_1(x,y) = -\alpha\frac{y}{r^2},\;\;\; A_2(x,y)=\alpha\frac{x}{r^2},
\end{align}
where $r^2=x^2+y^2$ and the index $i(=1,2)$ of $A_i$ denotes the direction of the vector ($x$ and $y$ components).
The field strength describing the vortex or a point-like flux at the origin is
\begin{align}
F_{12}=\partial_1 A_2-\partial_2A_1=2\pi \alpha\delta(x)\delta(y).
\end{align}

The Dirac Hamiltonian with the polar coordinate $(r,\theta)$ is
\begin{align}
  H &=\sigma_3 \left(\sum_{i=1,2}\sigma_i (\partial_i -iA_i) +m \right)
  = \sigma_3 \left(\begin{array}{cc} m & e^{-i\theta} \left( \frac{\partial}{\partial r}-i \frac{1}{r} \frac{\partial}{\partial\theta} -\frac{\alpha}{r} \right) \\ e^{i\theta} \left( \frac{\partial}{\partial r}+i \frac{1}{r} \frac{\partial}{\partial \theta} +\frac{\alpha}{r} \right) & m \end{array}\right).
\end{align}
Noting that $H$ conserves the angular momentum $J=-i\frac{\partial}{\partial \theta}+ \frac{1}{2} \sigma_3$,
the general solution for $H\psi^{E,j} = E\psi^{E,j}$ at $r\neq 0$,
which exponentially decays at large $r$ is given by
\begin{align}
  \label{eq:naivesol}
  \psi^{E,j}(r,\theta) = C\left(\begin{array}{c}
(m+E)K_{j-\frac{1}{2}-\alpha} (\sqrt{m^2-E^2} r)e^{i(j-\frac{1}{2})\theta }\\ 
         \sqrt{m^2-E^2} K_{j+\frac{1}{2}-\alpha} (\sqrt{m^2-E^2} r)e^{i(j+\frac{1}{2})\theta }
\end{array}\right),
\end{align}
where $C$ is the normalization constant, and $j$ is an eigenvalue of $J$,
taking a half integer: $j\in \frac{1}{2}+\mathbb{Z}$.

In the $r=0$ limit, 
the Bessel function is expanded as
$
K_\nu(M r) \sim  O(1/r^{|\nu|}), 
$ 
and the solution $\psi^{E,j}(r,\theta)$ is normalizable only when $j=\alpha$\footnote{
Note that $\int_{r_1} dr r r^{-2\beta}$ is finite in the $r_1\to 0$ limit
only when $\beta\le 1/2$.}.
The  $U(1)$ flux which allows the bound state of the electron is thus quantized.

The energy $E$ must be also quantized since the wave function
is limited in a tiny region around $r=0$.
If we smear the delta function in a region $r\sim r_1$,
the energy will be discretized in units of  $1/r_1$.
From the condition $E^2<m^2$, the only surviving
energy value in the $r_1\to 0$ limit is $E = 0$.

To summarize, for the vortex with $\alpha=1/2 + n$ with $n\in \mathbb{Z}$,
the system can have a unique square-integrable mode localized at the vortex,
\begin{align}
  \psi^{E=0,(2n+1)/2}(r,\theta) = C\left(\begin{array}{c}
m\\ 
|m|e^{i\theta }
\end{array}\right)e^{in\theta}K_{\frac{1}{2}} (|m|r),
\end{align}
whose energy is zero.
It is important to note that this mode is a ``chiral'' eigenstate of
\begin{align}
  \sigma_{r} = \frac{x}{r}\sigma_1+\frac{y}{r}\sigma_2 =
  \left(\begin{array}{cc}0 & e^{-i\theta}\\ e^{i\theta}& 0\end{array}\right),
\end{align}
with the eigenvalue $\text{sign}(m)$.

The solution explains that the vortex can
capture an electron state without any energy loss,
with which we can interpret that the vortex is ``charged''.
However, it is not sufficient to fully explain the physics in detail.
First, the solution does not describe why this happens in the topological insulator
with $m<0$ but does not in the normal insulator ($m>0$).
One may impose a boundary condition $\sigma_{r}\psi =\text{sign}(m)\psi$ by
hand for the topological insulator and $\sigma_{r}\psi =-\text{sign}(m)\psi$
for the normal insulator, but the choice looks ad-hoc and
any relation to the $U(1)$ gauge flux is unclear.
Second, the solution is square-integrable but its radial derivative $\partial_r\psi$ is not.
This may be related to the singularity of the $U(1)$ gauge field
but it is difficult to identify the origin.
Third, the Witten effect suggests that
the charge that the defect gains is a half integer.
But the above solution does not have any information about the fractional charge.

\subsection{Wilson term  and smearing of the $U(1)$ flux}

In the standard perturbative computation in continuum theory,
we take the Dirac equation as it is in the classical field theory
but higher derivative terms are introduced through
the Pauli-Villars or heat-kernel regularization
to make loop integrals finite.
In lattice gauge theory, on the other hand, we must regularize the
Dirac operator with the so-called Wilson term.
The Wilson term corresponds to a covariant Laplacian,
which is needed even at the tree-level to avoid fermion doubling.

The Wilson term can be interpreted as
a correction from the Pauli-Villars(PV) field
in Eq.~(\ref{eq:ZPV}).
The ``regularized'' Dirac operator is given by
\begin{align}
  D_\text{reg} &= M_\text{PV}\frac{D+m}{D+M_\text{PV}} = D+m -\frac{1}{M_\text{PV}}D_\mu D^{\mu},
\end{align}
where $O(1/M^2_\text{PV},m/M_\text{PV},F_{\mu\nu}/M_{PV} )$ terms are omitted.
Note that the overall factor $M_\text{PV}$ is multiplied to
keep the standard normalization.
If we identify $1/M_\text{PV}$ as the lattice spacing,
the additional term corresponds to the Wilson term.

The Wilson term seems to play no role in the continuum limit
$M_\text{PV}\to  \infty$ because it monotonically vanishes.
However, as shown below, when the gauge field is
focused in a smaller range compared to $1/M_\text{PV}$,
it gives a nontrivial effect.
It is important to note that the sign of the mass $m$ is
well-defined once the sign of $M_\text{PV}$ is fixed.
The Dirac equation is manifestly different between positive and negative $m$.
As mentioned in the previous sections,
negative $m$ represents the topological insulator, while
$m>0$ is in the trivial phase.

In the following analysis, we consider the Dirac Hamiltonian
rather than the Dirac operator containing temporal derivative.
The Wilson term is introduced only for the spacial directions,
which corresponds to taking the continuum limit in the time direction first\footnote{
In our study where $A_0=0$ and $A_{1,2}$ are $t$-independent,
there is no contribution from the temporal covariant derivative.
}.

Let us also regularize the gauge field by 
considering the $U(1)$ flux with a finite size:
\begin{align}
A_1(x,y) &= -\alpha\frac{y}{r_1^2},\;\;\; A_2(x,y)=\alpha\frac{x}{r_1^2} \;\;\; (\mbox{for } r\le r_1),  \nonumber\\
A_1(x,y) &= -\alpha\frac{y}{r^2},\;\;\; A_2(x,y)=\alpha\frac{x}{r^2} \;\;\; (\mbox{for } r> r_1),
\end{align}
where we assume $r_1<1/M_\text{PV}$. The field strength $F_{12}$ is zero
outside of the circle $r> r_1$, while it is smeared to a constant $F_{12}=2\alpha/r_1^2$
inside $r\le r_1$, keeping the total flux
\begin{align}
\int_{r<r_1} d^2x F_{12} = 2\pi \alpha,
\end{align}
unchanged. In the polar coordinate, it is equivalent to setting the gauge field
in the angular direction by
\begin{align}
  A_\theta = \left\{\begin{array}{cc}
  \frac{\alpha r}{r_1^2} & (\mbox{for } r\le r_1)\\
\frac{\alpha}{r} & (\mbox{for } r> r_1)
  \end{array}\right..
\end{align}

Note that the vector potential is regular at $r=0$.
In the above set up, the covariant Laplacian is given by
\begin{align}
  D_i D^i = \frac{\partial^2}{\partial r^2}+\frac{1}{r}\frac{\partial}{\partial r} -\left( -i\frac{1}{r}\partial_\theta-A_\theta\right)^2.
\end{align}

\subsection{Solving the regularized Dirac equation}

We are now ready to solve the regularized Dirac equation
\begin{align}
  H_\text{reg}\psi
  &=
  \left[H-\frac{1}{M_\text{PV}}\sigma_3 D_iD^i \right]\psi
  = E\psi,
\end{align}
where we denote the energy by $E$.
Since the system still respects the rotational symmetry: 
the modified Dirac Hamiltonian commutes with
$J=-i\frac{\partial}{\partial \theta}+ \frac{1}{2} \sigma_3$,
we can take the eigenfunction in the form
\begin{align}
\psi^{E,j} = \left(\begin{array}{c}f(r)e^{i\theta(j-1/2)}\\g(r)e^{i\theta(j+1/2)}\end{array}\right),
\end{align}
where $j$ takes a half integer. On the $(f,g)$ space, $H_\text{reg}$ acts as
\begin{align}
  \label{eq:Heff}
  \left(\begin{array}{cc} m-\frac{1}{M_\text{PV}}
      \left[\frac{\partial^2}{\partial r^2}+\frac{1}{r}\frac{\partial}{\partial r}
        -\left( \frac{j-1/2}{r}-A_\theta\right)^2 \right]
      &
      \frac{\partial}{\partial r}+\left(\frac{j+1/2}{r} -A_\theta\right) \\
    - \frac{\partial}{\partial r}+\left(\frac{j-1/2}{r} -A_\theta \right) &
      -m+\frac{1}{M_\text{PV}}\left[\frac{\partial^2}{\partial r^2}+\frac{1}{r}\frac{\partial}{\partial r}
        -\left( \frac{j+1/2}{r}-A_\theta\right)^2 \right]\end{array}\right).
\end{align}

\subsubsection{Solution at $r>r_1$}

In the region $r>r_1$, $A_\theta=\alpha/r$.
Noting that the modified Bessel equation guarantees that $K_\nu(\kappa r)$ (as well as $I_\nu(\kappa r)$)
for any positive value of $\kappa$
is the eigenstates of the Laplacian:
\begin{align}
\left[\frac{\partial^2}{\partial r^2}+\frac{1}{r}\frac{\partial}{\partial r}
        - \frac{\nu^2}{r^2}\right]K_\nu(\kappa r) = \kappa^2 K_\nu(\kappa r),
\end{align}
the solution to the Dirac equation can still be in the form
\begin{align}
 \left(\begin{array}{c}a K_{j-1/2-\alpha}(\kappa r)\\b K_{j+1/2-\alpha}(\kappa r)\end{array}\right) 
\end{align}
with the coefficients $a$ and $b$.

The general solution is thus a linear combination of the two solutions
\begin{align}
\label{generalsolout}
  \psi_{\rm out}^{E,j}(r,\theta) &= C\left(\begin{array}{c}
\left(m+E-\frac{\kappa_-^2}{M_\text{PV}}\right)K_{j-\frac{1}{2}-\alpha} (\kappa_- r)e^{i(j-\frac{1}{2})\theta }\\ 
         \kappa_- K_{j+\frac{1}{2}-\alpha} (\kappa_- r)e^{i(j+\frac{1}{2})\theta }
  \end{array}\right)\nonumber\\&+
  D\left(\begin{array}{c}
\left(m+E-\frac{\kappa_+^2}{M_\text{PV}}\right)K_{j-\frac{1}{2}-\alpha} (\kappa_+ r)e^{i(j-\frac{1}{2})\theta }\\ 
         \kappa_+ K_{j+\frac{1}{2}-\alpha} (\kappa_+ r)e^{i(j+\frac{1}{2})\theta }
  \end{array}\right)
\end{align}
with the coefficients $C$ and $D$, where
\begin{align}
\kappa_\pm = M_\text{PV}\sqrt{\frac{(1+2m/M_\text{PV})\pm \sqrt{(1+2m/M_\text{PV})^2-4(m^2-E^2)/M_\text{PV}^2}}{2}}.
\end{align}
The other two solutions given by $I_\nu$ are not square integrable at $r\to \infty$.

\subsubsection{Solution at $r\le r_1$}

Next, we consider the $r\le r_1$ region.
In this case, $A_\theta = \alpha r/r_1^2$ and
the covariant Laplacian operates as
\begin{align}
  D_i D^i = \frac{\partial^2}{\partial r^2}+\frac{1}{r}\frac{\partial}{\partial r} -\frac{1}{r^2}
  \left[\left(j\mp \frac{1}{2}\right)-\frac{\alpha r^2}{r_1^2}\right]^2,   
\end{align}
to the $\sigma_3=\pm 1$ component.

Let us look at the upper spinor component $f(r)$.
Assuming $j>0$ and taking the following form
\begin{align}
f(r) = a r^{j-1/2}e^{-\frac{\alpha r^2}{2r_1^2}}F(r),
\end{align}
with a constant $a$, we find
with the change of the variable $t=\alpha r^2/r_1^2$
that
\begin{align}
  D_i D^i f(r) =& a r^{j-1/2}e^{-\frac{\alpha r^2}{2r_1^2}}
  \frac{4\alpha}{r_1^2}
  \left[t\frac{\partial^2}{\partial t^2}+\left\{\left(j+\frac{1}{2}\right)-t\right\}\frac{\partial}{\partial t}
    -\left(-\frac{r_1^2}{4\alpha}L\right)\right]F(r)\nonumber\\
  &+a r^{j-1/2}e^{-\frac{\alpha r^2}{2r_1^2}}\left(-L-\frac{2\alpha}{r_1^2}\right)F(r),
\end{align}
where we can take an arbitrary constant $L$ (which cancels out at the right-hand side).
If we take $F(r)$ to be a confluent hypergeometric function: $F_1(-r_1^2L/4\alpha,j+1/2; \alpha r^2/r_1^2 )$,
then the first term vanishes and $f(r)$ is an eigenstate of the Laplacian:
\begin{align}
  D_i D^i f(r) = \left(-L-\frac{2\alpha}{r_1^2}\right)f(r).
\end{align}

Taking the lower spinor component as
\begin{align}
g(r) = b\frac{L}{2j+1} r^{j+1/2}e^{-\frac{\alpha r^2}{2r_1^2}}{}_1F_1(-r_1^2L/4\alpha+1,j+3/2; \alpha r^2/r_1^2 ),
\end{align}
the Dirac equation is in a closed form\footnote{
In the computations, we have used the following formulas
about the confluent hypergeometric functions (here we omit the subscripts).
\begin{align}
  \partial_z F(\alpha,\gamma;z)&=\frac{\alpha}{\gamma} F(\alpha+1,\gamma+1;z),
  \\
  z\partial_z F(\alpha,\gamma+1;z)&=\gamma\left[ F(\alpha,\gamma;z)- F(\alpha,\gamma+1;z)\right],
  \\
  z F(\alpha+1,\gamma+1;z)&=(1-\gamma/\alpha) F(\alpha,\gamma+1;z)
  +\frac{\gamma}{\alpha} F(\alpha,\gamma;z),
  \\
  (\alpha-\gamma) F(\alpha,\gamma+1;z) &=\alpha F(\alpha+1,\gamma+1;z)-\gamma F(\alpha,\gamma;z).
\end{align}  
} with respect to the coefficients 
$(a,b)$: 
\begin{align}
\left(\begin{array}{cc} m+\frac{1}{M_\text{PV}}
      \left(L+\frac{2\alpha}{r_1^2}\right)
      & L
       \\
    1 &
    -m-\frac{1}{M_\text{PV}}\left(L-\frac{2\alpha}{r_1^2}\right)
\end{array}\right)\left(\begin{array}{c}
a\\ 
b
\end{array}\right) = E\left(\begin{array}{c}
a\\ 
b
\end{array}\right).
\end{align}
Note that the same equation is obtained 
for negative $j$.
The general solution is thus a linear combination of the two solutions
\begin{align}
\label{generalsolin}
  \psi_{\rm in}^{E,j}(r,\theta) &= A\left(\begin{array}{c}
\left(m+E+\frac{L_-}{M_\text{PV}}-\frac{2\alpha}{M_\text{PV}r_1^2}\right)f^j_-(r)e^{i(j-\frac{1}{2})\theta }\\ 
         g^j_-(r)e^{i(j+\frac{1}{2})\theta }
  \end{array}\right)\nonumber\\&+
  B\left(\begin{array}{c}
\left(m+E+\frac{L_+}{M_\text{PV}}-\frac{2\alpha}{M_\text{PV}r_1^2}\right)f^j_+(r)e^{i(j-\frac{1}{2})\theta }\\ 
         g^j_+(r)e^{i(j+\frac{1}{2})\theta }
  \end{array}\right),
\end{align}
with constant coefficients $A$ and $B$, where
\begin{align}
  L_\pm = M^2_\text{PV}
  \left[\frac{-(1+2m/M_\text{PV})\pm \sqrt{(1+2m/M_\text{PV})^2
        -4\left\{m^2-(E-\frac{2\alpha}{M_\text{PV}r_1^2})^2\right\}/M_\text{PV}^2}}{2}\right],
\end{align}
and
\begin{align}
  f^j_\pm(r)&= \left\{
  \begin{array}{ll}
    r^{j-1/2}e^{-\frac{\alpha r^2}{2r_1^2}}{}_1F_1(-r_1^2L_\pm/4\alpha,j+1/2; \alpha r^2/r_1^2 ) & (\mbox{for } j>0)\\
    r^{-j+1/2}e^{-\frac{\alpha r^2}{2r_1^2}}{}_1F_1(-r_1^2L_\pm/4\alpha-j+1/2,-j+3/2; \alpha r^2/r_1^2 ) &  (\mbox{for } j<0)
  \end{array}\right.,\\
  g^j_\pm(r)&= \left\{
  \begin{array}{ll}
    \frac{L_\pm}{2j+1}r^{j+1/2}e^{-\frac{\alpha r^2}{2r_1^2}}{}_1F_1(-r_1^2L_\pm/4\alpha+1,j+3/2; \alpha r^2/r_1^2 ) & (\mbox{for } j>0)\\
    (2j-1)r^{-j-1/2}e^{-\frac{\alpha r^2}{2r_1^2}}{}_1F_1(-r_1^2L_\pm/4\alpha-j+1/2,-j+1/2; \alpha r^2/r_1^2 ) &  (\mbox{for } j<0)
  \end{array}\right.,
\end{align}

The other two solutions given by Tricomi's hypergeometric function $U(a,b,z)$
are not square integrable.

\subsubsection{Connecting in and out}

Let us connect the exact solutions at $r=r_1$.
In the standard analysis, one requires the eigenfunction and its first derivative
to be continuous.
In the above case where both solutions are eigenfunctions of the Laplacian,
it is equivalent to imposing the two conditions
\begin{align}
\psi_{\rm in}^{E,j}(r_1,\theta)=\psi_{\rm out}^{E,j}(r_1,\theta), \;\;\;D_i D^i\psi_{\rm in}^{E,j}(r_1,\theta)=D_i D^i\psi_{\rm out}^{E,j}(r_1,\theta),
\end{align}
together with the normalization condition
to determine the coefficients $A,B,C,D$ and the energy eigenvalue $E$.

In the limit $|M_{\rm PV}|\ll 1/r_1$,
we can analytically show the existence of a deep bound state
near $E\sim 0$.
To this end, let us first rewrite the eigenfunction as
\begin{align}
  \psi_{\rm in}^{E,j}(r_1,\theta)  = A \psi_{\rm in}^- + B\psi_{\rm in}^+, \;\;\;
  \psi_{\rm out}^{E,j}(r_1,\theta)  = C \psi_{\rm out}^- + D\psi_{\rm out}^+,
\end{align}
and that with the Laplacian multiplied accordingly.
The explicit form for each component is given in Eqs.~(\ref{generalsolout}) and (\ref{generalsolin}).
In order for the coefficients $A,B,C,D$ to have nontrivial
solutions, the energy $E$ must satisfy
\begin{align}
  \label{eq:detABCD}
  \det \left(\begin{array}{cccc}
\psi_{\rm in}^- & \psi_{\rm in}^+ & -\psi_{\rm out}^- & -\psi_{\rm out}^+ \\
D_i D^i\psi_{\rm in}^- & D_i D^i\psi_{\rm in}^+ & -D_i D^i\psi_{\rm out}^- & -D_i D^i\psi_{\rm out}^+ \\
  \end{array}\right) = 0.
\end{align}

In the $|M_{\rm PV}|\ll 1/r_1$ limit\footnote{A simpler limit $m\ll |M_{\rm PV}|\ll 1/r_1$ was investigated in \cite{Aoki:2022aez} by two of the authors.} where 
$L_\pm\sim  \pm \frac{2\alpha}{r_1^2}-\frac{M_{\rm PV}(M_{\rm PV}+2m)}{2}$,
a term containing the lower component of 
$D_i D^i\psi_{\rm in}^-$ 
and the upper component of 
$D_i D^i\psi_{\rm in}^+$ 
gives the dominant contribution to the determinant.
Therefore, Eq.~(\ref{eq:detABCD}) reduces to
\begin{align}
  \label{eq:reddetABCD}
  \det \left(\begin{array}{cc}
 -\psi_{\rm out}^- & -\psi_{\rm out}^+\\ 
  \end{array}\right) = 0,
\end{align}
which requires $\psi_{\rm out}^-$ and $\psi_{\rm out}^+$ to be parallel to each other.

It is important to note that when $\alpha=j$, 
the Dirac operator becomes real 
: $\sigma_3H^*=(\sigma_1e^{-i(2j\theta)})^{-1}\sigma_3H(\sigma_1e^{-i(2j\theta)})$. 
For the one-dimensional edge modes, this real structure is identified as 
the $T$ symmetry.
Since $H^* = -(\sigma_1e^{-i(2j\theta)})^{-1}H(\sigma_1e^{-i(2j\theta)})$, any eigenmode of $H$ with a non-zero eigenvalue $\lambda$:
$H\psi_\lambda=\lambda\psi_\lambda$, makes a pair with $\sigma_1e^{-i(2j\theta)}\psi_\lambda^*$,
whose eigenvalue is  $-\lambda$.
Therefore, if the eigenvalue below $|m|$ is unique,
$E=0$ is the only possible choice, to keep the spectrum $\pm$ symmetric.

Let us explicitly confirm the above argument taking by the $r_1\to 0$ limit and $\alpha-[\alpha]=(1+\epsilon)/2$,
while keeping $m$ and $M_{\rm PV}$ arbitrary under the condition $mM_{\rm PV}<0$.
The condition Eq.~(\ref{eq:reddetABCD}) leads to
\begin{align}
\frac{\kappa_+^{1+\epsilon}-\kappa_-^{1+\epsilon}}{\kappa_+^{-1+\epsilon}-\kappa_-^{-1+\epsilon}}=M_{\rm PV}(m-E).
\end{align}
In the small $\epsilon$ expansion, the energy is given by
\begin{align}
E = \epsilon\frac{2|m|}{\sqrt{1+4m/M_{\rm PV}}}\ln\left[\frac{1+\sqrt{1+4m/M_{\rm PV}}}{\sqrt{1+4m/M_{\rm PV}}}\right],
\end{align}
which becomes zero in the $\epsilon\to 0$ limit.

In the same $r_1\to 0$ limit with finite $M_{\rm PV}$ and $m$, setting $\alpha=j$ (and $E=0$),
let us also compute the coefficients $A,B,C,$ and $D$.
It is sufficient to consider the angle $\theta=0$.
In this limit, we have $\kappa_\pm=M_{\rm PV}(1\pm\sqrt{1+4m/M_{\rm PV}})/2$ (where we have assumed $m<0$)
and $(m-\kappa_\pm^2/M)=-\kappa_\pm$.
From the connection condition for $D_i D^i\psi_{\rm in/out}^{E=0,j}(r_1,\theta=0)$,
we obtain 
\begin{align}
  A = r_1^{3-j}A'\frac{(j+1/2)e^{j/2}}{(2j)^2{}_1F_1(3/2,j+3/2;j)},\;\;\;
  B = r_1^{2-j}A'\frac{e^{j/2}}{jM_{\rm PV}{}_1F_1(-1/2,j+1/2;j)},
\end{align}
with 
\begin{align}
  A' = -\sqrt{\frac{\pi}{2}}(C\kappa_-^{5/2}+D\kappa_+^{5/2}).
\end{align}

It is interesting to note that we can immediately conclude $\psi_{\rm in}^{E=0,j}(r_1,\theta)=0$
in the limit $r_1=0$ with the coefficients $A$ and $B$ above.
Then we finally obtain from the condition $0= \sqrt{r_1}\psi_{\rm out}^{E=0,j}(r_1,\theta=0)$
that $D=-\sqrt{\kappa_-/\kappa_+}C$ and the bound state's eigenfunction 
\begin{align}
\label{exactzeromode}
 \psi_{\rm out}^{E=0,j}(r,\theta) &=
  C'\left[\sqrt{\kappa_-}K_{\frac{1}{2}}(\kappa_-r)-\sqrt{\kappa_+}K_{\frac{1}{2}}(\kappa_+ r)\right]
  \left(\begin{array}{c}e^{i(j-1/2)\theta}\\-e^{i(j+1/2)\theta}\end{array}\right),
\end{align}
where $C'=-C\sqrt{\kappa_-}$ is a dimensionless normalization constant,
which agrees with the solution obtained by Ref.\cite{Mesaros:2012hu}.

Note in the large $M_\text{PV}$ limit,
$\kappa_+\to M_\text{PV}$, while $\kappa_-\to |m|$.
Therefore, the $\kappa_+$ component decays quickly 
and the only $\kappa_-$ component survives  
converging to  Eq.~(\ref{eq:naivesol}).


\subsection{Microscopic understanding and why the charge is fractional}

We have obtained an exact solution of the
regularized Dirac equation with the Wilson term
with a finite radius $r_1$ of the $U(1)$ flux.
Here we give a microscopic interpretation of the results.

The Wilson term makes the sign of the fermion mass well-defined.
Since every eigenvalue of the covariant Laplacian $-D_i D^i$ is positive
for any configuration of the gauge field,
inside topological insulators or in the case $m<0$ taking $M_{\rm PV}$ positive,
it is possible to locally flip the sign of the ``effective'' mass of the fermion
\begin{align}
m_{\rm eff} = \left\langle m+\frac{-D_i D^i}{M_{\rm PV}}\right\rangle,
\end{align}
by a strong magnetic flux.
In Ref.~\cite{Aoki:2022aez}, we numerically confirmed
that the vortex makes a small island of the positive mass region.

With the $U(1)$ flux analyzed above, the solution at $r<r_1$ has
a dominant component whose eigenvalue of $-D_i D^i$ is $O(1/r_1)$.
Therefore, for any value of $m<0$, the $r_1\to 0$ limit
guarantees creation of a domain-wall around $r=r_1$ on which
a single chiral edge-localized mode appears, while the wall never appears
inside normal insulators with $m>0$.
This is the unique bound state that the vortex captures
in the topological insulators.
The sharp change of the mass can be identified 
as the origin of the steep $r$ dependence of the wave function.

Moreover, when $\alpha=\frac{1}{2}$ mod $\mathbb{Z}$, the
Dirac operator becomes real in the $r_1\to 0$ limit.
Then the binding energy must be zero to make the
spectrum $\pm$ symmetric.
For the one-dimensional edge-localized effective Hamiltonian,
this is the $T$-symmetric point where the $T$
transformation is given by the complex conjugate operation.
For this real Dirac operator, the number of zero modes
is a topological invariant known as the mod-two Atiyah-Singer index  \cite{AtiyahSinger1971TheIndex5}.

In order to discuss the topological feature of the fermion zero mode,
we also need an infra-red regularization of the whole system.
So far we have assumed an infinite flat space $\mathbb{R}^2$.
In particle physics, the system is often regularized by
the one-point compactification identifying the infinite points $r=\infty$.
However, this does not work in our case.
If we succeeded in the one-point compactification,
the topological insulator $m_{\rm eff}<0$ region would have
a topology of a disk with a small $S^1$ boundary at $r=r_1$.
This contradicts the fact that the mod-two AS index is
a cobordism invariant, which cannot be nonzero
if the target manifold is a boundary of some higher-dimensional manifold.
The resolution is given by setting $m>0$ region outside to create
another domain-wall at, say, $r=r_0$.
In physics, this treatment is obvious since for any topological insulator,
there exists its outside in the normal phase\footnote{Another consistent
  treatment is to put the same number of vortices 
and anti-vortices with even number of domain-walls in the system.}.

Once the new domain-wall is set, the cobordism requires
another zero mode at $r=r_0$ to keep the total index trivial.
Then the two zero eigenmodes, the bound state at the vortex
and the one at the new domain-wall, mix by
tunneling effect and the eigenvalues are split to some symmetric
small values $\pm \varepsilon$
(still keeping the $T$ symmetry of the edges). 
The edge-localized modes on a domain-wall with radius $r_0\gg r_1$
were analytically obtained in Ref.~\cite{Aoki:2022aez} by two of the authors.
The zero energy mode at $\alpha=j$ is a chiral eigenstate
with $\sigma_r=+1$, which is opposite to that the solution
in Eq.~(\ref{exactzeromode}) has.

The 50\% 
of the mixed state 
is located at the vortex, while the other 50\%
 is sitting at the domain-wall.
 In Ref.~\cite{Aoki:2022aez}, two of the authors numerically confirmed
 this splitting on a square lattice.
 For the half-filling situation where we put the Fermi energy to zero,
 only one of the two split zero modes is occupied.
 Then the charge expectation value around the vortex is $-1/2$ of the unit charge,
 while the other $-1/2$ is distributed on the surface of the
 topological insulator.

 This is our microscopic interpretation of the mechanism
 how the vortex in the two-dimensional topological insulator
 gains a half of the electric charge.
 As shown below, the same mechanism explains the Witten effect whereby
 a magnetic monopole in a three-dimensional topological insulator
 becomes a dyon with a half-integral electric charge.

\section{A monopole in three dimensions}
\label{sec:monopole}

Next let us consider a magnetic monopole in three dimensions.
The vector potential of the Dirac monopole is given by
\begin{align}
A_1=\frac{-q_m y}{r(r+z)},\;\;\;A_2=\frac{q_m x}{r(r+z)},\;\;\;A_3=0,
\end{align}  
(the subscript $3$ denotes the $z$-direction) of which
field strength is 
\begin{align}
  \partial_i A_j- \partial_j A_i= q_m \epsilon_{ijk}\frac{x_k}{r^3} - 4\pi q_m\delta(x)\delta(y)\Theta(-z)\epsilon_{ij3},
\end{align}
where we denote $(x_1,x_2,x_3)=(x,y,z)$.
The second term represents the so-called Dirac string with the step function $\Theta(-z)=+1$ for $z\le 0$, and $\Theta(-z)=0$, otherwise.
In order to eliminate the Dirac string, we may introduce another
vector potential, which is regular in the thin cone around
the Dirac string and glues to the first one, from which
the quantization condition of the magnetic charge is required.
However, in this work, we regard the Dirac string as a physical object
between the monopole and anti-monopole at a long distance.
Here and in the following, we assume $q_m=n/2$ with an integer $n$.

When the Dirac string has no physical effect, the monopole configuration has
an $SO(3)$ rotational symmetry.
Therefore, it is convenient to introduce the orbital angular momentum operator
\begin{align}
L_i=-i\epsilon_{ijk} x_j (\partial_k -iA_k)-n \frac{x_i}{2r},
\end{align}  
which satisfies
$[L_i,L_j]=i\epsilon_{ijk} L_k$.
Their explicit form in the polar coordinate is given by 
\begin{align}
    L_{\pm}=&L_1\pm iL_2 =e^{ \pm i\phi } \left(\pm \frac{\partial}{\partial \theta} +i \frac{\cos \theta}{ \sin \theta} \frac{\partial}{\partial \phi} +\frac{n}{2} \frac{\cos \theta -1}{ \sin \theta} \right), \\
    L_3=& -i\frac{\partial}{\partial \phi} -\frac{n}{2}.
\end{align}

The highest and lowest states of $L_3$ are obtained from the equation $L_\pm g_\pm=0$ as
\begin{align}
    g_+=&\sin^l \theta \left( \frac{1-\cos\theta}{1+\cos\theta}\right)^\frac{n}{4} e^{i(l+\frac{n}{2}) \theta}=(\sin \theta e^{i\phi})^{l+\frac{n}{2}} (1+\cos \theta)^{-\frac{n}{2}},  \\
    g_-=&\sin^l \theta \left( \frac{1+\cos\theta}{1-\cos\theta}\right)^\frac{n}{4} e^{i(l+\frac{n}{2}) \theta}=(\sin \theta e^{-i\phi})^{l-\frac{n}{2}} (1+\cos \theta)^{\frac{n}{2}},
\end{align}
for which the two conditions $l \pm \frac{n}{2} \in \mathbb{Z}$
from the periodicity of $\phi$, 
and
$l- |\frac{n}{2}|\geq 0$ required from regularity at $\theta\to 0$,
must be satisfied.
Consequently, $l$ must be a half integer not smaller than $|\frac{n}{2}|$,
which is in contrast to the free fermion case with an integer $l$.

\subsection{Naive Dirac equation}

With the monopole background obtained above, 
let us review the study of a bound state 
of the naive Dirac Hamiltonian \cite{Yamagishi:1982wp}
without the Wilson term,
\begin{align}
  H &=\gamma_0 \left(\gamma^i (\partial_i -iA_i) +m \right)
  = \left(\begin{array}{cc} m & \sigma^i\left(\partial_i-iA_i \right) \\
    -\sigma^i\left(\partial_i-iA_i \right) & -m \end{array}\right),
\end{align}
where we have set $\gamma_0=\sigma_3\otimes 1$ and $\gamma_i=\sigma_1\otimes \sigma_i$.
It is important to note that $H$ anti-commutes with a ``chirality''\footnote{
$\bar\gamma$ is different from
the standard chirality operator $\gamma_5=-\gamma_0\gamma_1\gamma_2\gamma_3=\sigma_2\otimes 1$.
The extra symmetry with $\bar\gamma$ comes from absence of the scalar potential.
} operator $\bar\gamma=\sigma_1\otimes 1$.

The orbital angular momentum operator alone does not commute with
the Dirac Hamiltonian but the total angular momentum
\begin{align}
J_i = L_i +\frac{1}{2}\sigma_i,\;\;\;[J_i,J_j]=i\epsilon_{ijk} J_k,
\end{align}
does: $[1\otimes J_i, H]=0$ follows from  $[J_i, \sigma^j (\partial_j -iA_j ) ]=0$.
 Except for the lowest eigenvalue $j=|\frac{n}{2}|-1/2$, 
where the degeneracy is $2j+1=|n|$,
there are $2(2j+1)$ degenerate states.
In the following analysis, we use with $\sigma_r=x^j \sigma_j/r$ that
\begin{align}
[J_i, \sigma_r]=0, \;\;\;
\sigma^i L_i = J^2-L^2-3/4.
\end{align}

There is another operator which commutes with $H$.
To find this, let us define
\begin{align}
 D^{S^2}:= \sigma^i\left(L_i+\frac{n}{2}\frac{x_i}{r}\right)+1
\end{align}
whose physical meaning will be discussed later.
Using the following equalities,
\begin{align}
 [J_i,D^{S^2}]=0, \;\;\;
 \{D^{S^2},\sigma_r\}=0, \;\;\;
  \left[D^{S^2}\right]^2=
  \left(j+\frac{1}{2}\right)^2-\frac{n^2}{4}.
\end{align}
it is not difficult to show $[\sigma_3\otimes D^{S^2}, H]=0$.

For  $j>|\frac{n}{2}|-1/2$ let us introduce the eigenstates of $D^{S^2}$
which satisfies
\begin{align}
D^{S^2}\chi_{j,j_3,\pm}(\theta,\phi)&=\pm \sqrt{\left(j+\frac{1}{2}\right)^2 -\frac{n^2}{4} }\chi_{j,j_3,\pm}(\theta,\phi).
\end{align}
Then for any function $f(r)$  and $g(r)$, the vector of the form
\begin{align}
  \label{eq:solform}
  \frac{1}{\sqrt{r}}\left(
  \begin{array}{c}f(r)\chi_{j,j_3,\pm}(\theta,\phi)\\
    g(r)\sigma_r\chi_{j,j_3,\pm}(\theta,\phi)\end{array}\right),
\end{align}
is an eigenstate of $\sigma_3\otimes D^{S^2}$.
Here the overall factor $1/\sqrt{r}$ is introduced just for later convenience.

The  $j=|\frac{n}{2}|-1/2$ state we denote by $\chi_{j,j_3,0}$
is special and does not make $\pm$ pair of $D^{S^2}$. 
We note that $\chi_{j,j_3,0}$ is an eigenstate of $\sigma_r$:
\begin{align}
\sigma_r \chi_{j,j_3,0}(\theta,\phi) = {\rm sign}(n)\chi_{j,j_3,0}(\theta,\phi).
\end{align}  
Since the degeneracy at each $j$ is exhausted, the $\theta$ and $\phi$
dependence is completely obtained.

The remaining $r$ dependence determines the energy eigenvalue $E$ of $H$.
For $j>|\frac{n}{2}|-1/2$, we find no normalizable solution 
localized at the monopole.
The only candidate for  the bound state that the monopole can capture is, therefore,
the state with $j=|\frac{n}{2}|-1/2$, which should have the form
\begin{align}
  \label{eq:solform0}
  \frac{1}{r}\left(
  \begin{array}{c}f(r)\chi_{j,j_3,0}(\theta,\phi)\\
    g(r)\chi_{j,j_3,0}(\theta,\phi)\end{array}\right).
\end{align}
and the equation reduces to
\begin{align}
  \left(\begin{array}{cc} m& {\rm sign}(n)\partial_r \\
    -{\rm sign}(n)\partial_r & -m \end{array}\right)\left(
  \begin{array}{c}f(r)\\
    g(r)\end{array}\right)=E\left(
  \begin{array}{c}f(r)\\
    g(r)\end{array}\right).
\end{align}
Noting that $H^2$ operates diagonally and identically on $f$ and $g$,
we can deduce that $g=\pm f$, or equivalently $\bar\gamma=\sigma_1\otimes 1=\pm 1$  and $E=0$.
The solution is $f(r)=\exp(-|m|r)$ and the possible bound state is, thus,
\begin{align}
  \label{eq:naiveedge}
    \frac{C_{j,j_3,0}}{r}\exp(-|m|r)\left(
  \begin{array}{c}1\\ {\rm sign}(m){\rm sign}(n)\end{array}
    \right)\chi_{j,j_3,0}(\theta,\phi),
\end{align}
which is a chiral eigenstate of $\sigma_1\otimes \sigma_r$ with
the eigenvalue ${\rm sign}(m)$. Here the normalization constant is denoted by $C_{j,j_3,0}$.

For a unit magnetic charge $n=1$, $j=j_3=0$ the
possible bound state above is unique.
However, we cannot distinguish normal and topological
insulators unless we specify the chiral boundary condition at $r=0$.
In the above analysis, the origin of the chirality is not clear, either.
In Ref.~\cite{Yamagishi:1982wp}, a half electric charge is obtained
by summing up all the charges in the Dirac sea and comparing with that without monopole,
but only after a regularization which breaks conservation of the charge.

\subsection{Wilson term}

Let us introduce the Wilson term
to the Dirac Hamiltonian in the presence of a monopole,
\begin{align}
  H &=\gamma_0 \left(\gamma^i (\partial_i -iA_i) +m -\frac{D_iD^i}{M_{\rm PV}}\right)
  = \left(\begin{array}{cc} m -D_iD^i/M_{\rm PV}& \sigma^i\left(\partial_i-iA_i \right) \\
    -\sigma^i\left(\partial_i-iA_i \right) & -m +D_iD^i/M_{\rm PV}\end{array}\right).
\end{align}

First, we examine the symmetry of the modified Dirac Hamiltonian.
Using 
\begin{align}
  \left[\sigma^i (\partial_i -iA_i)\right]^2
  = D_iD^i + \frac{n}{2r^2}\sigma_r,
\end{align}  
$[J_i,\sigma_r]=0$, and $[J_i, \sigma^j (\partial_j -iA_j ) ]=0$
we can show that $[D_iD^i, J_i]=0$ and the rotational symmetry is maintained: $[H, 1\otimes J_i]=0$.

However,
\begin{align}
  \left[D^{S^2}, D_iD^i\right]
\neq 0,
\end{align}
which indicates that the $2(2j+1)$ degenerate states for $j> |n/2|-1/2$
are split to two $(2j+1)$ degenerate sets.

It is remarkable that $\bar\gamma = \sigma_1\otimes 1$ still anti-commutes with $H$,
which indicates that if the bound state around the monopole is unique,
its energy eigenvalue $E$ must be zero and it must be a chiral eigenstate of $\bar\gamma$.

Before the exact calculation, let us determine the
chirality of the bound state.
Let us assume that $M_{\rm PV}>0$, $m<0$ and the point-like magnetic
monopole charge is smeared in the region $r<r_1$.
In condensed-matter setup, the PV mass $M_\text{PV}$ corresponds
to the lattice spacing around 10$^{-10}$ $\rm{m}$ between the atoms, 
while the size of the monopole is less than 10$^{-20}$ $\rm{m}$, assuming the monopole energy 
where the theory of 't~Hooft--Polyakov \cite{tHooft:1974kcl,Polyakov:1974ek} applies is higher than $10$ TeV.
From a dimensional analysis, $-D_iD^i\sim 1/r_1^2$ and under the condition $r_1\ll 1/M_\text{PV}$,
it is natural to assume that
$m_{\rm eff}(r) = \langle m-D_iD^i/M_{\rm PV}\rangle$ is effectively positive for some region near the origin,
while it is kept negative at large $r$.
Namely, a domain-wall is created around the monopole.
Then the bound state should satisfy\footnote{
Here we treat the effect of the Wilson term as a mean-field for the Dirac equation.
See the exact treatment in the next subsection.
}
\begin{align}
\left[  \sigma_1\otimes\sigma_r \partial_r +m_{\rm eff}(r)\right]\psi = 0,
\end{align}
to which the normalizable solution is given by
\begin{align}
\psi \sim \exp\left[\int_{r_1}^{r}dr'm_{\rm eff}(r')  \right],
\end{align}
which requires $\sigma_1\otimes \sigma_r =-1$.
It is interesting to note that the both chiralities $\bar\gamma =\pm 1$
are possible as long as the lower-dimensional chirality
takes the same eigenvalue $-1\otimes \sigma_r =\pm 1$.
This reflects the fact that the Atiyah-Singer index on the two-dimensional
domain-wall is not $\mathbb{Z}_2$ but $\mathbb{Z}$.
In fact, we will show below that the index is equal to  the magnetic charge $n$.


\subsection{The solution to the regularized Dirac equation}

Let us derive the bound state described by the regularized Dirac equation.
Here we assume that $j= |n/2|-1/2$, the edge-localized states have a chirality 
$\sigma_1\otimes \sigma_r =-1$ at $r_1$, and take the $r_1\to 0$ limit.

We employ the ansatz for the (nearest) zero modes
to have the following form with $s={\rm sign}(n)$
\begin{align}
  \label{eq:solform0W}
  \psi=\frac{f(r)}{r}\left(
  \begin{array}{c}1\\
    -s\end{array}\right)\otimes \chi_{j,j_3,0}(\theta,\phi),
\end{align}
which has $2j+1=|n|$ degeneracy, having different values of $j_3$.
It is important to note that
$\psi$ is a simultaneous eigenstate of $-\bar\gamma=-\sigma_1\otimes 1$ and $1\otimes \sigma_r$
sharing the same eigenvalue $s$.
From the anti-commutation relation $\{H,\bar\gamma\}=0$,
one can conclude that the eigenvalue of $H$ is exactly zero,
as long as it is isolated.

With a further change of the functional form to $f(r)=\sqrt{r}e^{-M_{\rm PV}r/2}g(\kappa r)$ 
with $\kappa=\frac{M_{\rm PV}}{2}\sqrt{1+4m/M_{\rm PV}}$
it is not difficult to show that $g(\kappa r)$ satisfies the modified Bessel equation.
Here we have assumed that $4|m|/M_{\rm PV}<1$.
Therefore, we obtain the general solution as
\begin{align}
  \psi=\frac{e^{-\frac{M_{\rm PV}r}{2}}}{\sqrt{r}}
\left[AK_\nu(\kappa r)+BI_\nu(\kappa r)\right]
  \left(
  \begin{array}{c}1\\
    -s\end{array}\right)\otimes \chi_{j,j_3,0}(\theta,\phi),
\end{align}
where $\nu=(\sqrt{2|n|+1})/2$, with the constant coefficients $A$ and $B$.

In order to connect $\psi$ and $D_i D^i\psi\sim \psi/r$ (indicated by the Bessel equation)
at finite $r=r_1$ and take the $r_1\to 0$ limit, the
steep $1/r_1$ dependence requires (in the same way as explicitly shown in the two-dimensional vortex case in Sec.~\ref{sec:vortex})
that $A=O(r_1^\nu)$ ($\psi$ vanishes at $r_1\to 0$).

\begin{figure}[tbhp]
\begin{center}
  \includegraphics[width=8cm]{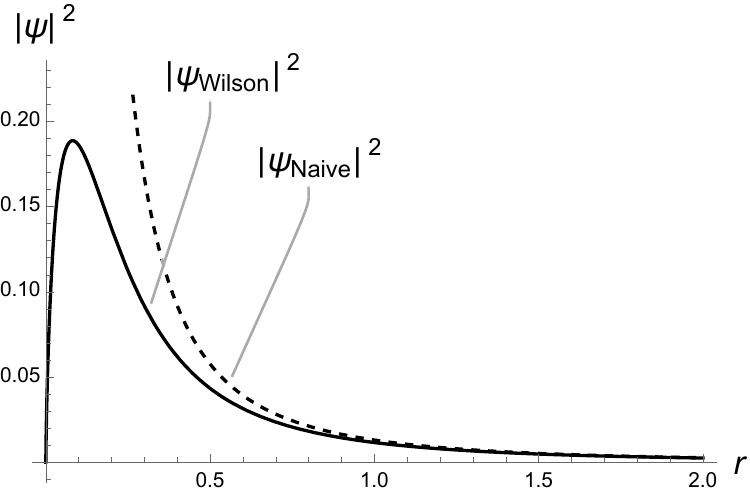}
\end{center}
\caption{
The comparison of the solution with the Wilson term in Eq.~(\ref{monosol})
(solid curve)
and the one in the previous work (dashed curve) in Eq.~(\ref{eq:naiveedge}).
Here we choose the parameters $n=1, m=0.1$ and $M_\text{PV}=10$.
}
\label{Fig:NaiveVSWilson}
\end{figure}

We finally obtain the localized state to the monopole\footnote{
Our solution coincides the result obtained in Refs.\cite{zhao2012magnetic,Tyner:2022qpr}
where  the Wilson term is introduced but smearing of the point-like singularity of the monopole
and dynamical mass shift were not considered.
},
\begin{align}
 \label{monosol}
  \psi^\text{mono}_{j,j_3}=\frac{Be^{-\frac{M_{\rm PV}r}{2}}}{\sqrt{r}}
I_\nu(\kappa r)
  \left(
  \begin{array}{c}1\\
    -s\end{array}\right)\otimes \chi_{j,j_3,0}(\theta,\phi),
\end{align}
with a normalization constant $B$.
As is expected, in the $M_{\rm PV}\to \infty$ limit,
where $I_\nu(\kappa r)\sim \exp\left(\frac{M_{\rm PV}}{2}r-|m|r\right)/\sqrt{\pi r M_{\rm PV}}$,
$\psi$ converges to
the naive Dirac solution in Eq.~(\ref{eq:naiveedge}).
The comparison of the two results is shown in Fig.~\ref{Fig:NaiveVSWilson}
where we choose $n=1, m=0.1$ and $M_\text{PV}=10$.
Remarkable differences is that the Wilson term yields 
a peak at $r=c_\nu/M_\text{PV}$ with $c_\nu=(4\nu^2-1)/4=|n|/2$ 
and makes the wave function zero at $r=0$.

In contrast to the vortex system in Sec.~\ref{sec:vortex}
the magnetic field is nonzero even outside $r_1$.
In fact, as shown in Fig.~\ref{Fig:NaiveVSWilson} the peak of the edge mode wave function in Eq.~(\ref{monosol})
is located at $r\sim 1/M_\text{PV}$.
The scale $1/M_\text{PV}$ corresponds to the atomic lattice spacing $\sim 10^{-10}$ m of the topological insulator,
which will be much larger than the size of the monopole $\sim 10^{-20}$ m.
We conclude by this exact solution that the domain-wall with a finite radius $r\sim 1/M_\text{PV}$
is created by the monopole, on which the $|n|$ chiral edge modes appear
with energy zero. These are the origin of the electric charge the magnetic monopole gains.

\subsection{Atiyah-Singer index theorem and zero mode pairing}

In the previous subsection, we have obtained
$|n|=2j+1$ zero modes with the chirality $1\otimes \sigma_r=s=\mbox{sign}(n)$.
The fact that these zero modes are located not on the point-singularity of the monopole
but on a finite two-dimensional spherical domain-wall
makes the topological property of the solutions clearer.
Note that they are also the zero modes of the operator $D^{S^2}=\sigma^i\left(L_i+\frac{n}{2}\frac{x_i}{r}\right)+1$.
In fact, this operator can be identified as
the effective Dirac operator on the two-dimensional sphere around the monopole 
with a radius we denote by $r_2\sim 1/M_\text{PV}$.
With a local Lorentz (or $Spin^c$ to be precise) transformation
$R(\theta,\phi)=\exp(i\theta\sigma_2/2)\exp(i\phi(\sigma_3+1)/2)$,
we obtain
\begin{align}
  \frac{1}{r_2}D^{S^2} &:= \frac{1}{r_2}R(\theta,\phi)D^{S^2}R(\theta,\phi)^{-1}
=- \frac{1}{r_2}\sigma_3\left[\sigma_1\frac{\partial}{\partial \theta}
    +\sigma_2\left(\frac{1}{\sin\theta}\frac{\partial}{\partial \phi}+i\hat A_\phi +i\hat A^s_\phi\right)\right],
\end{align}
where $\hat A_\phi=\frac{n}{2}\frac{\sin\theta}{1+\cos\theta}$ 
is the vector potential (in units of $r_2$)
generated by the monopole, and
\begin{align}
\hat A^s_\phi = \frac{1}{2\sin\theta}-\frac{\cos\theta}{2\sin\theta}\sigma_3,
\end{align}
is the induced $Spin^c$ connection
on the sphere which  is strongly curved with the small radius $r_2$ \cite{Aoki:2022aez}.
It is also important to note that the same transformation
changes the chirality operator to $R(\theta,\phi)\sigma_rR(\theta,\phi)^{-1}=\sigma_3$.
Namely, these zero modes are the chiral zero modes of $D^{S^2}$.

As shown in Ref.~\cite{Aoki:2022aez}, the gravitational effect on a sphere generally 
gives a gap in the massless Dirac operator spectrum with a size of its inverse radius.
The $U(1)$ gauge connection from the monopole cancels 
the gravity to have the zero eigenvalues.
Since the degeneracy of the zero modes is $|n|$ with the chirality $\sigma_3=\mbox{sign}(n)$,
we can conclude that the index of the Dirac operator $D^{S^2}$ is $n$.

Stability of the chiral zero modes is topologically protected by the AS index theorem.
On the two-dimensional sphere of radius $r_2$, we can explicitly compute that
\begin{align}
\frac{1}{4\pi}\int_{S^2}d^2x \epsilon^{\mu\nu}F_{\mu\nu} = n.
\end{align}
Note that the $Spin^c$ connection $\hat A^s_\phi$ does not contribute to the index in two dimensions.

Since the AS index is a cobordism invariant,
a similar discussion about the zero mode pairing 
as in the case of the vortex in two dimensions in Sec.~\ref{sec:vortex}
applies here.
The conventional one-point compactification of $r=\infty$ points does not work,
since the  magnetic flux coming from the monopole becomes infinitely dense again at the compactified point,
which is equivalent to putting another monopole with the opposite magnetic charge $q_m=-n/2$.
This second monopole creates another domain-wall, which is cobordant to 
the one around the original monopole.
The argument implies that there must exist a region of normal phase 
before reaching $r=\infty$: outside of the topological insulator is always a normal insulator.
On the outer domain-wall, $|n|$ zero modes with 
the same chirality of $1\otimes \sigma_r$ as the monopole-localized zero modes must appear
as a consequence of the cobordism invariance of the AS index\footnote{
The cobordism invariance of the AS index can be understood by the Stokes theorem.
Suppose $\partial X_1$ and $-\partial X_2$
(where the negative sign indicates the opposite orientation to $X_1$)
 are two disjoint boundaries of a
three-dimensional manifold $X$. In the differential form, we can show
$0=\int_X dF = \int_{\partial X_1}F - \int_{-\partial X_2}F$.
}.

Let us consider another  domain-wall at large radius $r_0$, 
giving a position-dependent mass term
$m(r)=|m|\epsilon(r-r_0)$ where
$\epsilon(r-r_0)=\pm 1$ for $r \gtrless r_0$.
The region $r>r_0$ corresponds to a normal insulator.
The edge-localized state in each region is obtained as
\begin{align}
 \label{DWsol}
  \psi^\text{DW}_{j,j_3}=\left\{
  \begin{array}{cc}
  	\frac{\exp{\left({M_{\rm PV}r \over 2}\right)}}{\sqrt{r}}\left(e^{\kappa_- r_0}B' K_\nu(\kappa_- r)+e^{-\kappa_- r_0}C'  I_\nu(\kappa_- r) \right)
  \left(
  \begin{array}{c}1\\
    s\end{array}\right)\otimes \chi_{j,j_3,0}(\theta,\phi), & (r<r_0),\\
\frac{D'\exp\left(\kappa_+ r_0+\frac{M_{\rm PV}r}{2}\right)}{\sqrt{r}}
K_\nu(\kappa_+ r)
  \left(
  \begin{array}{c}1\\
    s\end{array}\right)\otimes \chi_{j,j_3,0}(\theta,\phi),& (r>r_0),
    \end{array}\right.
\end{align}
where  $\kappa_\pm=\frac{M_{\rm PV}}{2}\sqrt{1\pm 4|m|/M_{\rm PV}}$.
The ratios $C'/B'$ and $D'/B'$ are determined from
the connecting condition of the wave function, as well as that for its derivative at the domain-wall $r=r_0$.
In the large $r_0 m$ limit, we find
\begin{align}
	{C' \over B'}&=\frac{\kappa_-\left(K_{\nu-1}(\kappa_- r_0)+K_{\nu+1}(\kappa_- r_0)\right)K_\nu(\kappa_+ r_0)-\kappa_+\left(K_{\nu-1}(\kappa_+ r_0)+K_{\nu+1}(\kappa_+ r_0)\right)K_\nu(\kappa_- r_0)}{\kappa_-\left(I_{\nu-1}(\kappa_- r_0)+I_{\nu+1}(\kappa_- r_0)\right)K_\nu(\kappa_+ r_0)+\kappa_+\left(K_{\nu-1}(\kappa_+ r_0)+K_{\nu+1}(\kappa_+ r_0)\right)I_\nu(\kappa_- r_0)}e^{2\kappa_- r_0}  \nonumber\\
	&\sim \pi\frac{\kappa_- -\kappa_+}{\kappa_- + \kappa_+}, 
\end{align}
and 
\begin{align}
	{D' \over B'}&=\frac{\kappa_-\left\{\left(I_{\nu-1}(\kappa_- r_0)+I_{\nu+1}(\kappa_- r_0)\right)K_\nu(\kappa_- r_0)+\left(K_{\nu-1}(\kappa_- r_0)+K_{\nu+1}(\kappa_- r_0)\right)I_\nu(\kappa_- r_0)\right\}}{\kappa_-\left(I_{\nu-1}(\kappa_- r_0)+I_{\nu+1}(\kappa_- r_0)\right)K_\nu(\kappa_+ r_0)+\kappa_+\left(K_{\nu-1}(\kappa_+ r_0)+K_{\nu+1}(\kappa_+ r_0)\right)I_\nu(\kappa_- r_0)}e^{(\kappa_--\kappa_+)r_0} \nonumber\\
	&\sim \frac{2\sqrt{\kappa_- \kappa_+ }}{\kappa_- + \kappa_+}. 
\end{align}
These edge-localized modes have
the same $|n|=2j+1$ degeneracy with the zero modes captured by the monopole
but with the opposite chirality: $\sigma_1\otimes \sigma_r=+1$.


In the $M_\text{PV}\gg m$ limit, the zero mode above takes
a simpler form with $D'/B'\sim 1$ and $C'/B'\sim 0$:
\begin{align}
\psi^\text{DW}_{j,j_3}= \frac{B''}{r}\exp[-|m||r-r_0|]\left(
  \begin{array}{c}1\\
    s\end{array}\right)\otimes \chi_{j,j_3,0}(\theta,\phi).
\end{align}  

At finite $r_0$, the paired zero mode (having the same $j_z$)
near the monopole at $r_1$ and the domain-wall $r_0$ mix
to give symmetric splitting to the Dirac spectrum.
Let us perturbatively estimate this mixing following 
the method given in Ref.~\cite{zhao2012magnetic}.
When $|m| r_0$ is large enough, the zero mode around the monopole
$\psi^\text{mono}_{j,j_3}$ in Eq.~(\ref{monosol}) and
that near the domain-wall $\psi^\text{DW}_{j,j_3}$ in Eq.~(\ref{DWsol})
are still approximate eigenstates of $H$.
As the total angular momentum $J_i$ commutes with $H$,
each eigenstate of $H$ after mixing the two 
is given by 
\begin{align}
\psi = \alpha \psi^\text{mono}_{j,j_3} +\beta \psi^\text{DW}_{j,j_3},
\end{align}
where $\alpha$ and $\beta$ are constants.
Note here that $\psi^\text{mono}_{j,j_3}$ and $\psi^\text{DW}_{j,j_3}$
are both eigenstates of the chirality operator $\bar\gamma=\sigma_1\otimes 1$,
which anti-commutes with $H$.  We can immediately obtain the diagonal parts
$(\psi^\text{mono}_{j,j_3})^\dagger H \psi^\text{mono}_{j,j_3}=
(\psi^\text{DW}_{j,j_3})^\dagger H \psi^\text{DW}_{j,j_3}=0$.
We can also show that the off-diagonal part is real and symmetric:
$(\psi^\text{mono}_{j,j_3})^\dagger H \psi^\text{DW}_{j,j_3}=
(\psi^\text{DW}_{j,j_3})^\dagger H \psi^\text{mono}_{j,j_3}=:\Delta$.
Then we can show that $\alpha=\pm \beta$ and the split energy
is $E =\pm \Delta$.
This result, having 50\% amplitude around the monopole, 
is valid at any large value of $r_0$.
The normal insulator outside and the edge modes on it
cannot be ignored from the theory no matter how long they are separated.

When we set the Fermi energy zero, only one of each zero mode pairs is occupied.
Since only a half of the amplitude is distributed around the monopole,
the charge expectation value around $r=r_1$ will be $-|n|/2$,
while the other half is spread on the surface at $r=r_0$.
Thus we complete the microscopic description
how the monopole gains a half electric charge.
In the next section, we nonperturbatively examine the
above scenario on a lattice taking $r_0$ finite.

\section{Numerical analysis of monopole on a lattice}
\label{sec:lattice}

In this section, we numerically investigate the monopole
system in a topological insulator with a lattice regularization.
In particular, we try to confirm creation of the domain-wall,
or the surface of a normal insulator island around the monopole.
We will also quantify the zero mode pairing
of the monopole-captured state and the edge-localized state at
the spherical domain-wall with finite radius $r=r_0$ 
and their tunneling effects,
as well as the electric charge expectation value around the monopole.

\subsection{Lattice setup}

On three-dimensional hyper-cubic lattices with size $L=23, 31$ and 47
(where the number of sites in each direction is 24, 32 and 48, respectively),
with open boundary conditions,
we put a monopole at $\bm{x}_m=(x_m,y_m,z_m)=(L/2,L/2,L/2)$
with a magnetic charge $n/2$.
We also put an anti-monopole at $\bm{x}_a=(x_a,y_a,z_a)=(L/2,L/2,1/2)$
with the opposite charge  $-n/2$
to ensure the Gauss law constraint in the continuum limit.
The continuum vector potential at $\bm{x}=(x,y,z)$ is then given by
\begin{align}
  A_1(\bm{x})&=q_m\left[\frac{-(y-y_m)}{|\bm{x}-\bm{x}_m|(|\bm{x}-\bm{x}_m|+(z-z_m))}-
    \frac{-(y-y_a)}{|\bm{x}-\bm{x}_a|(|\bm{x}-\bm{x}_a|+(z-z_a))}\right],\nonumber\\
  A_2(\bm{x})&=q_m\left[\frac{(x-x_m)}{|\bm{x}-\bm{x}_m|(|\bm{x}-\bm{x}_m|+(z-z_m))}
    -\frac{(x-x_a)}{|\bm{x}-\bm{x}_a|(|\bm{x}-\bm{x}_a|+(z-z_a))}\right],\nonumber\\
  A_3(\bm{x})&=0,
\end{align}
with $q_m=n/2$.
Note that the Dirac string extends from $\bm{x}_a$ to $\bm{x}_m$.

The link variables on the lattice is given by exponentiated
line integrals of the vector potential:
\begin{align}
U_j(\bm{x}) = \exp\left[i \int_{0}^{1}A_{j}(\bm{x}')dl \right],
\end{align}
where $\bm{x}=(x,y,z)$ having integer-valued components
denotes the lattice coordinate,
and the line integral with respect to $\bm{x}'=\bm{x}+\bm{e}_j l$
where $\bm{e}_j$ is the unit vector in the $j$ direction
is taken from $\bm{x}$ to $\bm{x}+\bm{e}_j$ along the link.
Here and in the following, we take the lattice spacing to be unity.

\begin{figure}[tbhp]
\begin{center}
  \includegraphics[width=8cm]{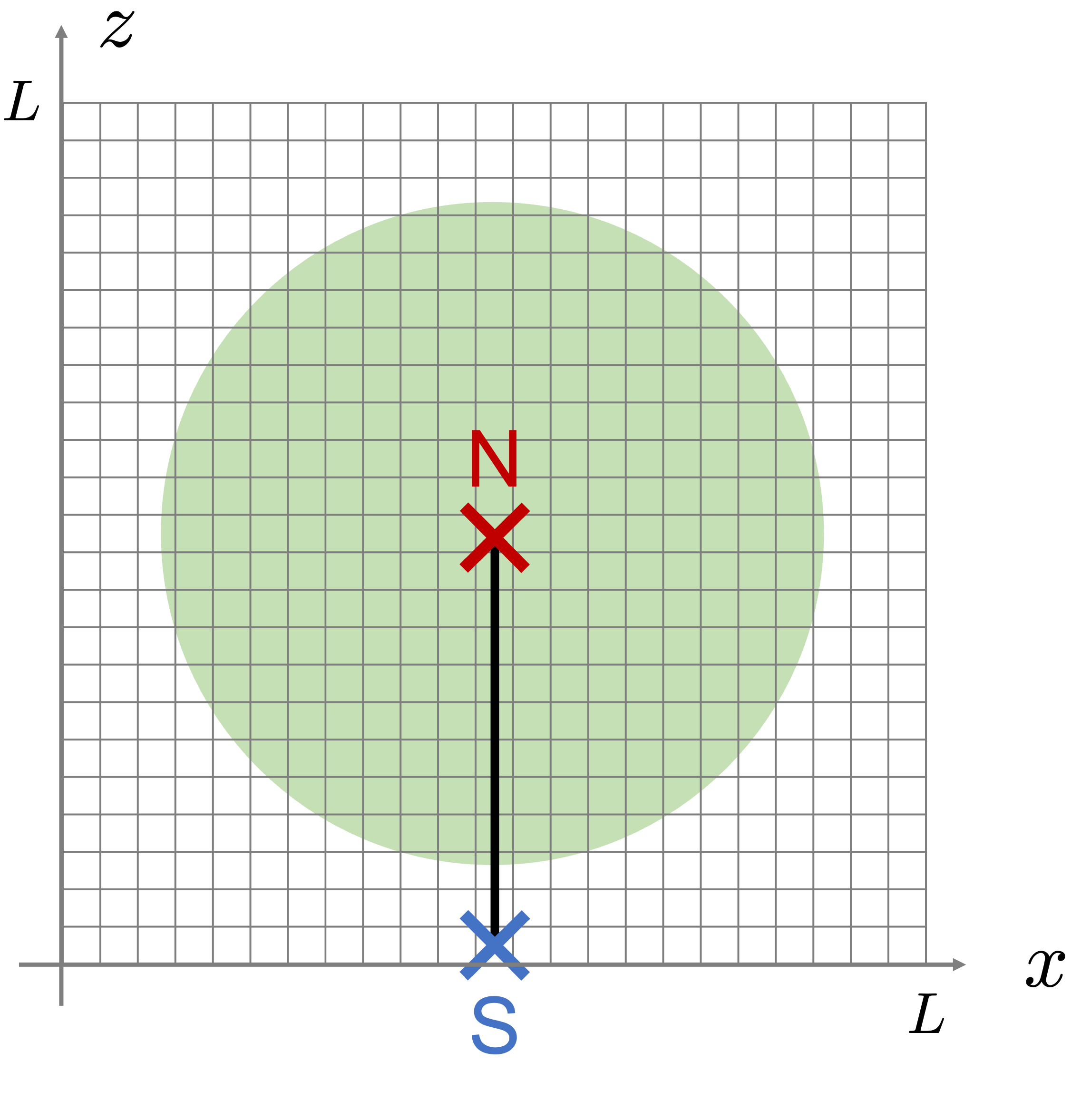}
\end{center}
\caption{Our lattice setup on a  $L=23$ lattice (at the $y=0$ slice) is presented.
The symbols ``N'' and ``S'' represent monopole and anti-monopole, respectively,
which are connected by the Dirac string shown by the thick line.
Inside the shadowed sphere of radius $r_0=9$, we put a negative mass $-m_0=-7/12$, while it is $+7/12$ outside.}
\label{Fig:setup}
\end{figure}

For the fermion field, 
we assign a position-dependent mass term $m(\bm{x})=-m_0$ 
for $r=\sqrt{|\bm{x}-\bm{x}_m|}\le r_0=\frac{3}{8} L$
and $m(\bm{x})=+m_0$ otherwise.
The constant $m_0 (L+1) =14$ is fixed to take the continuum limit. 
In this setup, the monopole is located at the center of
a spherical topological insulator with radius $r_0$,
while the anti-monopole sits in the normal insulator region with the mass $+m_0$.
Outside of the lattice with open boundary condition
corresponds to a ``laboratory'' with $m(\bm{x})=+\infty$.
Since the $m(\bm{x})=+m_0$ region and that with $m(\bm{x})=+\infty$
are both in the trivial phase, we do not expect any edge-localized modes
to appear at the boundaries $x_{i=1,2,3}=0$ or $x_{i=1,2,3}=L$.
Our lattice setup with $L=23$ is sketched in Fig.~\ref{Fig:setup}.
The symbols ``N'' and ``S'' represent monopole and anti-monopole, respectively,
which are connected by the Dirac string shown by the thick line.
Inside the spherical domain-wall of radius 
$r_0=9$, we put a negative mass $-m_0=-7/12$, while it is $+7/12$ outside.

The Wilson Dirac Hamiltonian is given by
\begin{align}
        H_W =\gamma^0 \left( \sum_{i=1}^3\left[\gamma^i\frac{\nabla^f_i+\nabla^b_i}{2} -\frac{1}{2}\nabla^f_i \nabla^b_i \right]+m(\bm{x}) \right), \label{eq:WilsonH}
     \end{align}
where $\nabla^f_i\psi(\bm{x})=U_i(\bm{x})\psi(\bm{x}+\bm{e}_i)-\psi(\bm{x})$
denotes the forward covariant difference and
$\nabla^b_i\psi(\bm{x})=\psi(\bm{x})-U^\dagger_i(\bm{x}-\bm{e}_i)\psi(\bm{x}-\bm{e}_i)$
is the backward difference.
The gamma matrices are the same as in the continuum:
$\gamma_0=\sigma_3\otimes 1$, $\gamma_i=\sigma_1\otimes \sigma_i$.
It is important to note that $H_W$ anti-commutes with $\bar\gamma=\sigma_1\otimes 1$ even on a lattice.
In the analysis below, we compute the eigenvalues denoted by $E_k$ and their eigenfunction $\phi_k(\bm{x})$
of $H_W$ and study their local amplitude normalized by $r^2$,
\begin{align}
A_k(\bm{x}) = \phi_k(\bm{x})^\dagger \phi_k(\bm{x})r^2,
\end{align}
as well as the local distribution of operators.

\subsection{Numerical results}

\begin{figure}[tbhp]
\begin{center}
  \includegraphics[width=8cm]{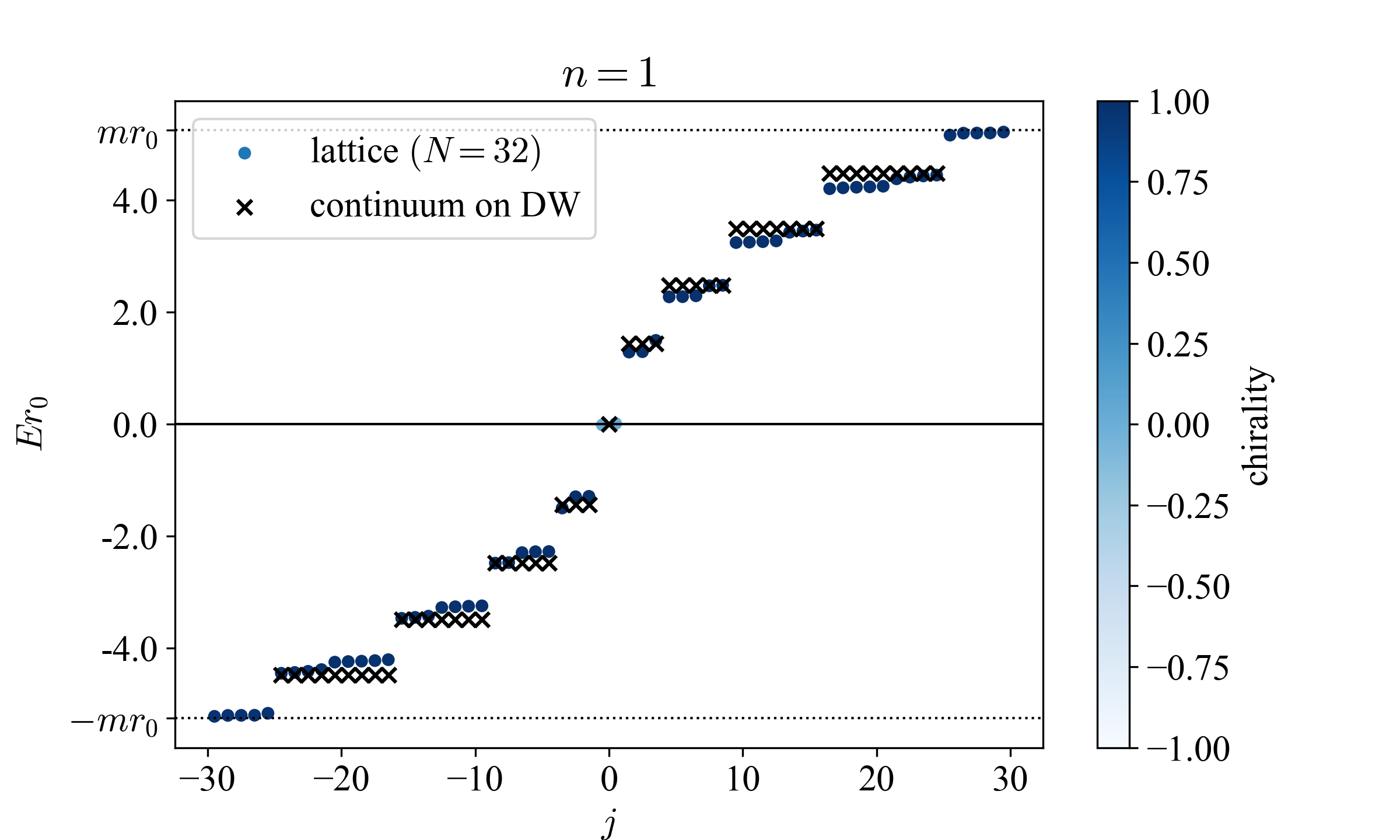}
 \includegraphics[width=8cm]{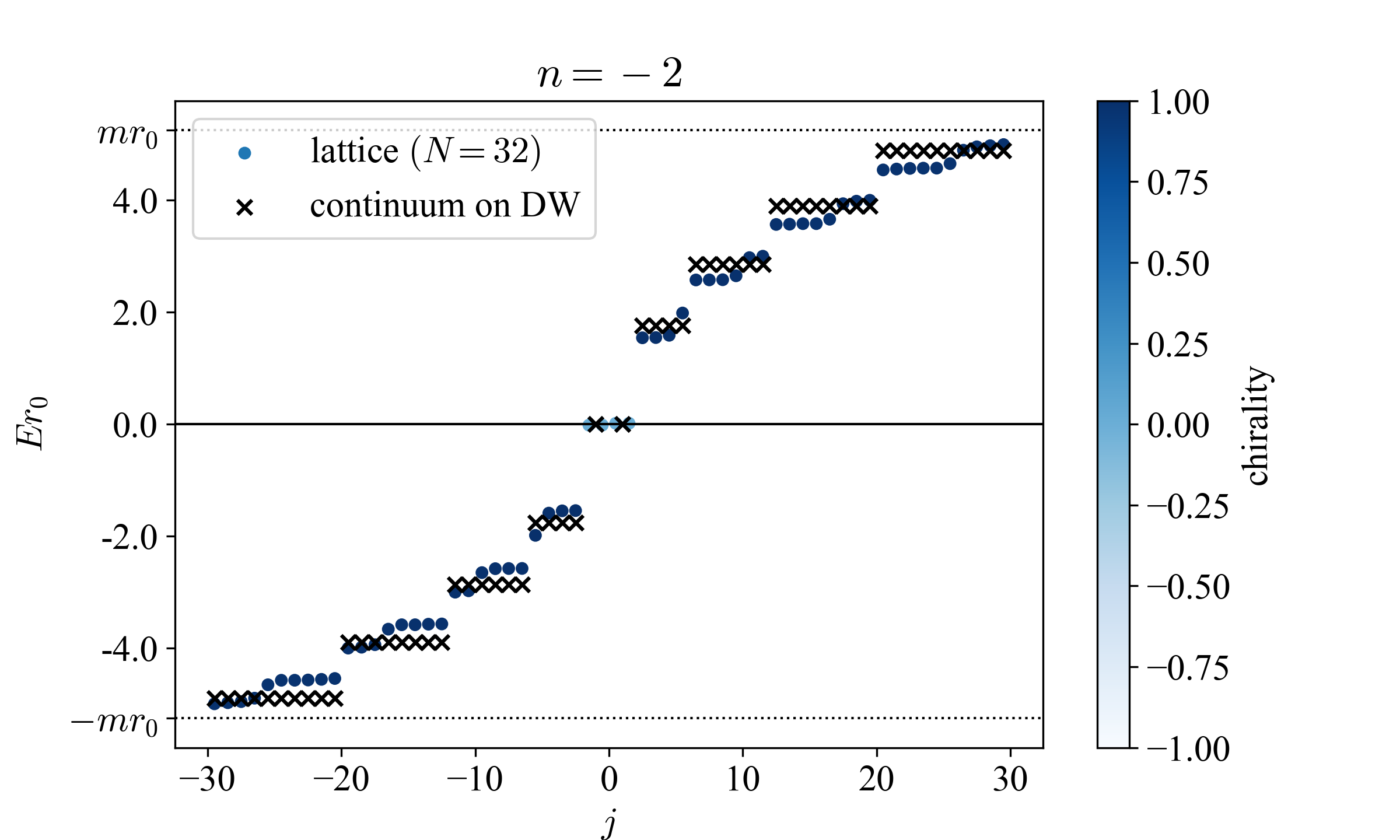}
\end{center}
\caption{Low-lying eigenvalue spectrum of $H_W$ with the monopole charge $n=1$ (left panel) and $n=-2$ (right).
  Circle symbols represent our lattice data, while crosses are continuum predictions for the edge-localized modes
  on the outer domain-wall.
  See the main text for details of other numerical setup.}
\label{Fig:spec}
\begin{center}
 \includegraphics[width=8cm]{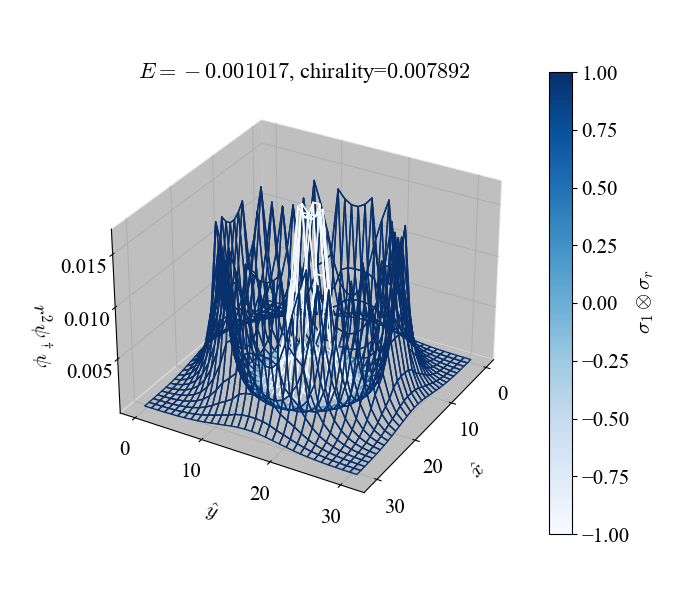}
 \includegraphics[width=8cm]{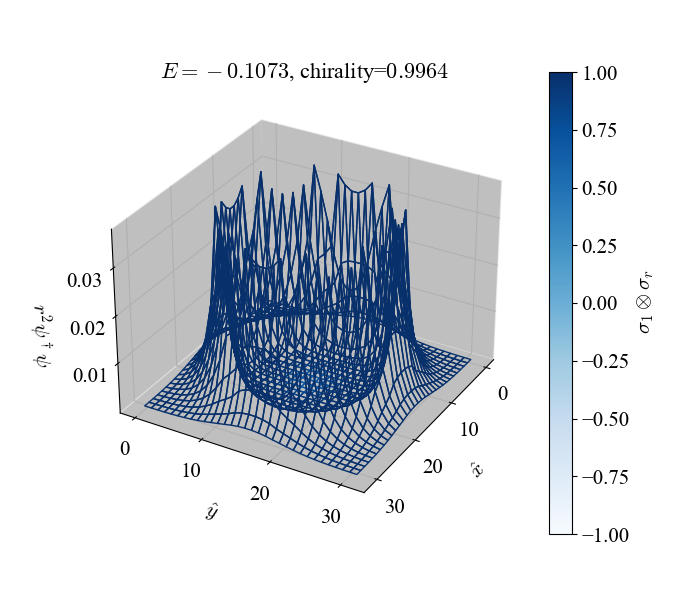}
\end{center}
\caption{Left panel: the amplitude of the negative nearest-zero mode $A_1(\bm{x})$ at the $z=(L+1)/2$ 
slice in the presence of a unit monopole $n=1$ is plotted. Right: the same as the left panel but for 
the second negative nearest-zero mode $A_2(\bm{x})$.
The gradation represents the locally measured chirality of $\sigma_1\otimes \sigma_r$.}
\label{Fig:A0}
\end{figure}

In Fig.~\ref{Fig:spec}, we plot the Dirac eigenvalue spectrum
on the $L=31$ lattice in units of $r_0$ with circle symbols.
On the left panel, the result with $n=1$ is presented
while that with $n=-2$ is shown in the right panel.
Our results are consistent with a similar numerical analysis presented in Ref.~\cite{Tyner:2022qpr}.
For comparison, we also plot the continuum results (without the Wilson term)
with crosses for the edge-localized modes on the domain-wall at $r_0$.
The gradation of the symbols for each eigen mode $\phi_k(\bm{x})$
represents the chirality expectation value measured  by
\begin{align}
\sum_{\bm{x}}\phi_k^\dagger(\bm{x})\left[\sigma_1\otimes \sigma_r\right]\phi_k(\bm{x}).
\end{align}
The edge-localized modes between 
$\pm m_0$ have chirality $\sigma_1\otimes \sigma_r \sim +1$,
except for the modes in the vicinity of zero.

Let us focus on the (near-)zero modes which are apparently not chiral.
The number of these modes is doubled compared to
the continuum prediction of the edge-localized modes around the domain-wall $r=r_0$.
As explained in the previous section, any chiral zero mode localized at the domain-wall
must appear in a pair with the mode with the opposite chirality localized at
the inner domain-wall dynamically created by the monopole.
The paired chiral zero modes are mixed due to the tunneling effect.
The near zero-mode doubling in our numerical data is consistent with this scenario.
For the negative nearest-zero mode in the case of $n=1$
(whose eigenfunction is denoted by $\phi_1(\bm{x})$)
we plot in the left panel of Fig.~\ref{Fig:A0} the amplitude $A_1(\bm{x})$
at the $z=(L+1)/2$ slice.
For comparison we also present the same plot in the right panel but for the second nearest-zero mode $A_2(\bm{x})$.
The gradation of the symbols shows the local chirality,
\begin{align}
\phi_k^\dagger(\bm{x})\left[\sigma_1\otimes \sigma_r\right]\phi_k(\bm{x})/\phi_k^\dagger(\bm{x})\phi_k(\bm{x}),
\end{align}
with $k=1$ or 2.
We can see only in the left panel that the amplitude has two peaks around $r=|\bm{x}-\bm{x}_m|=0$ 
and $r=r_0$ and the local chirality near each peak is $\sim -1$ and $+1$, respectively,
although the total chirality is close to zero.

\begin{figure}[tbhp]
\begin{center}
 \includegraphics[width=8cm]{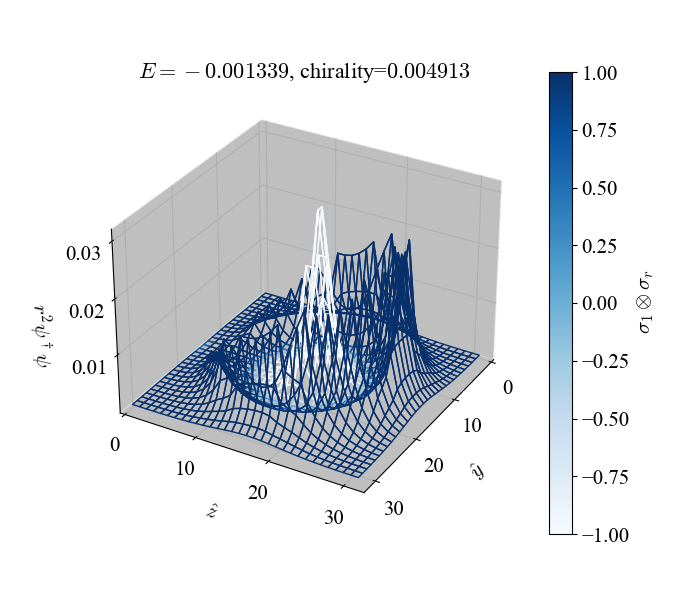}
 \includegraphics[width=8cm]{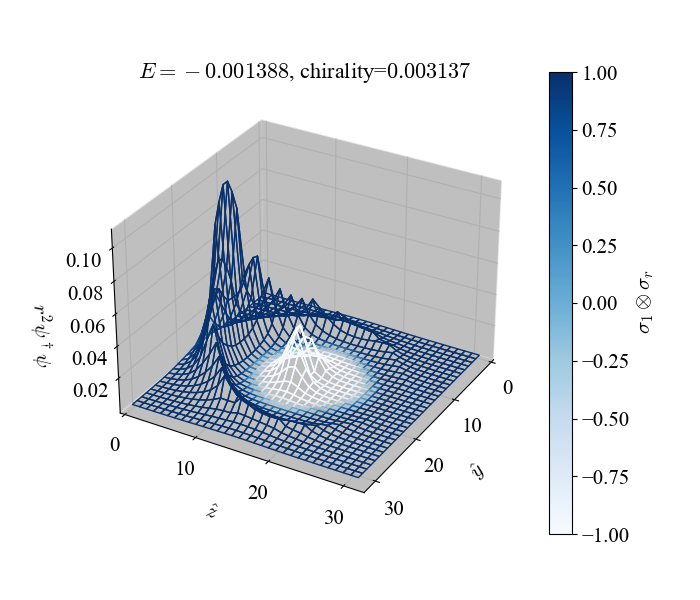}
\end{center}
\caption{Left panel: the amplitude of the negative nearest-zero mode $A_1(\bm{x})$ at the $x=(L+1)/2$ 
slice with monopole charge $n=-2$ is plotted. Right: the same as the left panel but for 
the second negative nearest-zero mode $A_2(\bm{x})$.}
\label{Fig:A0n-2}
\end{figure}

Figure~\ref{Fig:A0n-2} shows the same amplitudes
$A_0(\bm{x})$ (left panel) and  $A_2(\bm{x})$ (right)
but with $n=-2$ and at the $x=(L+1)/2$ slice.
As is expected, these are the two near-zero modes 
whose wave function is partly captured by the monopole.
The shape of the wave function becomes asymmetric
due to the anti-monopole located at $\bm{x}_a=(x_a,y_a,z_a)=(L/2,L/2,1/2)$, though.

\begin{figure}[tbhp]
\begin{center}
  \includegraphics[width=8cm]{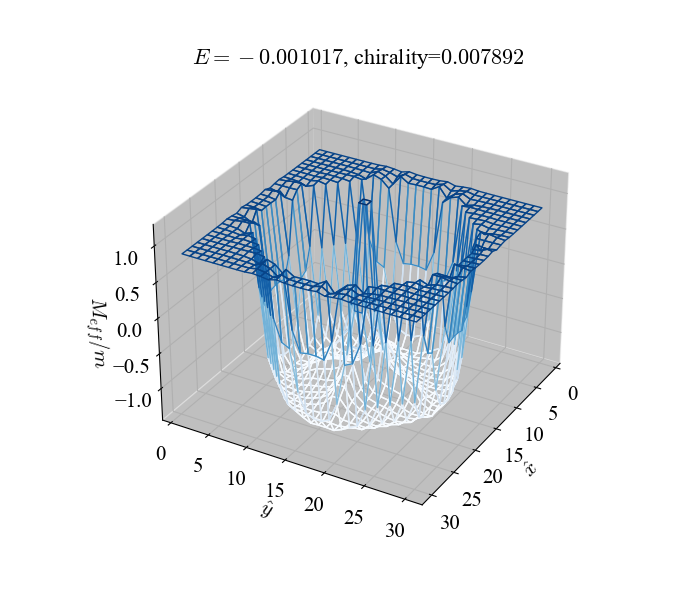}
  \includegraphics[width=8cm]{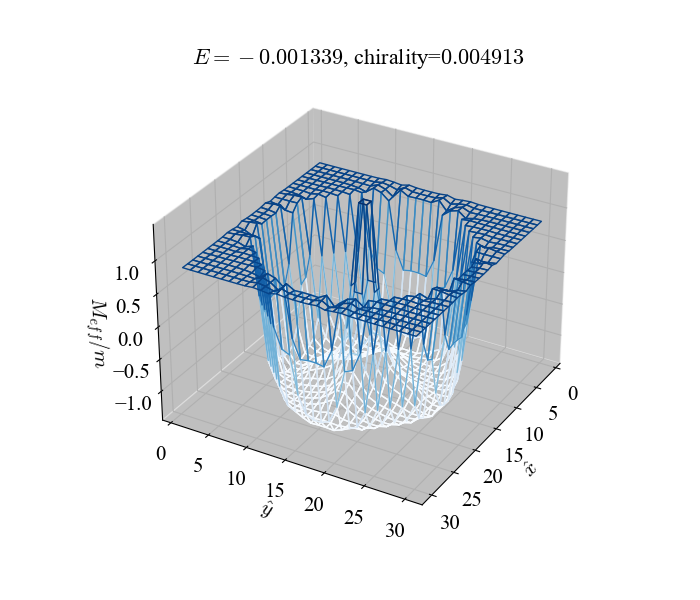}
\end{center}
\caption{The local effective mass (defined in the main text) normalized by $m_0$ at the $z=(L+1)/2$ slice
  for the negative nearest-zero mode $\phi_1$.
  The results for the magnetic charge $n=+1$ (left panel) and $n=-2$ (right) are plotted.
  The data are shown with gradation to make the two regions of normal phase with $m_\text{eff}/m_0\sim +1$ in dark mesh
  and the one for topological phase with $m_\text{eff}/m_0\sim -1$ in light mesh, explicit.
}
\label{Fig:massn1}
\end{figure}

Next let us directly confirm creation of the domain-wall near the monopole.
In Fig.~\ref{Fig:massn1} 
distribution of the ``effective mass''(normalized by $m_0$) 
locally measured at the $z=(L+1)/2$ slice,
\begin{align}
m_{\rm eff}(\bm{x}) = 
\left.\phi_1(\bm{x})^\dagger\left[-\sum_{i=1,2,3}\frac{1}{2}\nabla^f_i \nabla^b_i +m(\bm{x})\right]\phi_1(\bm{x})\right/\phi_1(\bm{x})^\dagger\phi_1(\bm{x}),
\end{align}
with a unit monopole charge $n=1$ (left panel)
and that 
with $n=-2$ (right) is plotted.
  The data are shown with gradation to make explicit the two regions of normal phase with $m_\text{eff}/m_0\sim +1$ in dark mesh
  and the one for topological phase with $m_\text{eff}/m_0\sim -1$ in light mesh.
We can see that a small island of a normal insulator
or a positive mass region appears around the monopole, 
and we thus confirm that a domain-wall is dynamically created.

\begin{figure}[tbhp]
\begin{center}
 \includegraphics[width=8cm]{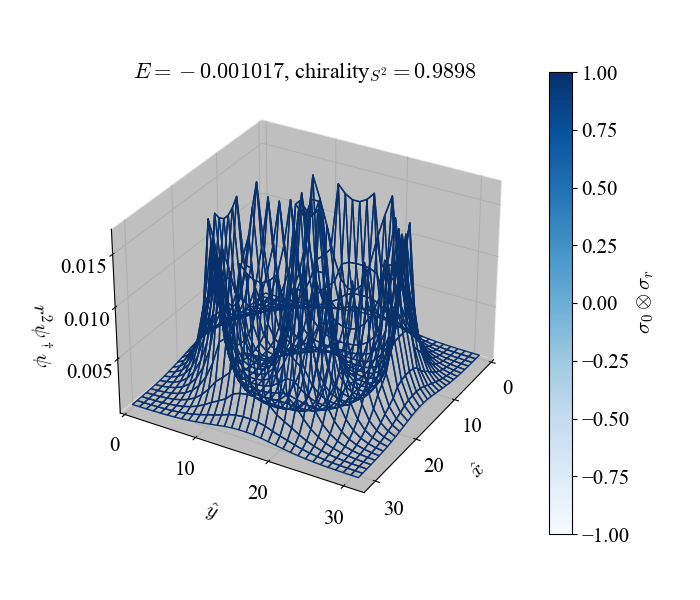}
 \includegraphics[width=8cm]{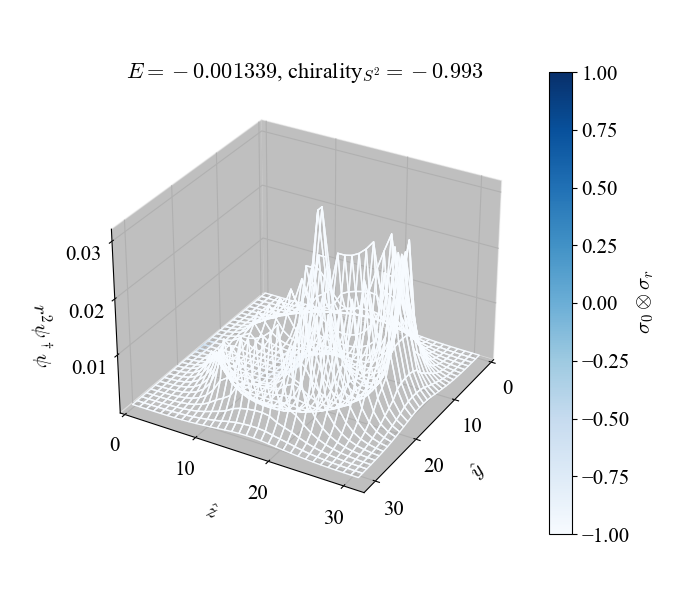}
\end{center}
\caption{The same plot as the left panel of Fig.~\ref{Fig:A0} and \ref{Fig:A0n-2} 
but with the gradation indicating the two-dimensional chirality $1\otimes \sigma_r$ locally measured.}
\label{Fig:A02Dchirality}
\end{figure}

As discussed above about Figs.~\ref{Fig:A0} and \ref{Fig:A0n-2}, 
the distribution of the chirality $\sigma_1\otimes \sigma_r$
has no significant difference between $n=1$  and $n=-2$ {\it i.e.},
the wave function near the monopole is $\sigma_1\otimes \sigma_r\sim -1$,
while that near the domain-wall at $r=r_0$ has $\sim +1$ chirality.
However, for the two-dimensional chiral operator $1\otimes \sigma_r$,
the result is sensitive to the sign of $n$.
As shown in Fig.~\ref{Fig:A02Dchirality} where 
the distribution of
\begin{align}
\phi_1^\dagger(\bm{x})\left[1\otimes \sigma_r\right]\phi_1(\bm{x})/\phi_1^\dagger(\bm{x})\phi_1(\bm{x})
\end{align}
is indicated by the gradation of the plot, the local two-dimensional chirality 
is uniformly $\sim +1$ for $n=1$
and $\sim -1$ for $n=-2$. 
This is consistent with the cobordism property of the AS index
that the spherical domain-wall at $r=r_0$
and that near the monopole must share the same value $n$.

\begin{figure}[tbhp]
\begin{center}
 \includegraphics[width=8cm]{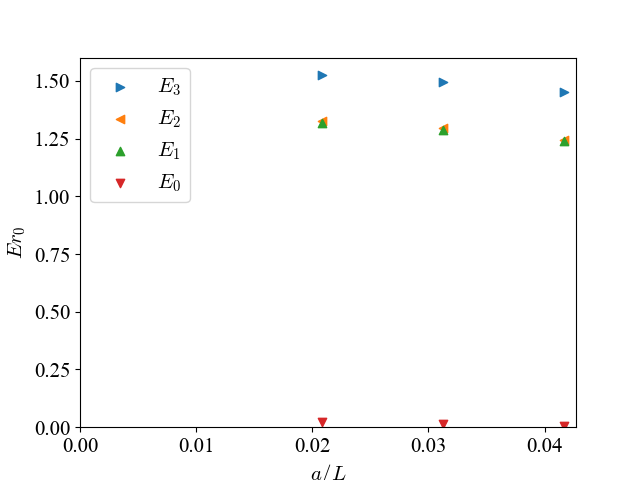}
 \includegraphics[width=8cm]{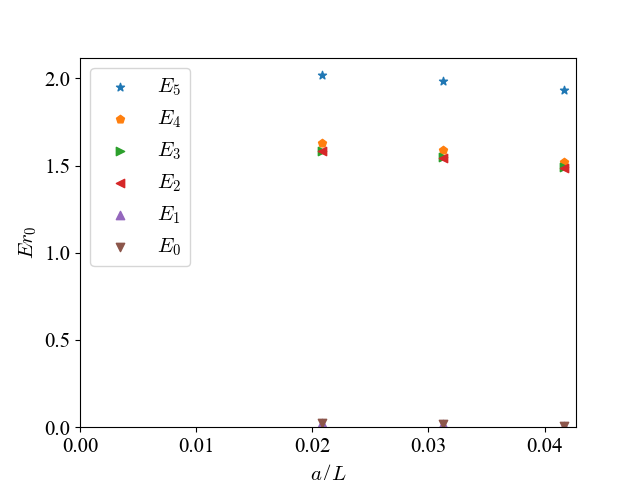}
\end{center}
\caption{The lattice cut-off $a/L$ dependence of the negative nearest-zero eigenvalues.
The results for $n=1$ (left panel) and those for $n=-2$ (right) are shown.}
\label{Fig:a-dep}
\end{figure}

So far we have used the lattice size $L=31$.
In order to estimate the systematics due to 
the lattice spacing $a$, we also compute the Dirac eigenvalues
on $L=23$ and $L=47$ lattices with the same physical set up.
In Fig.~\ref{Fig:a-dep} we present the lattice cut-off $a/L$ dependence 
of negative nearest-zero eigenvalues.
They show a mild linear dependence, which is comparable 
to what observed in the previous analysis in the free fermion case 
by two of the authors \cite{Aoki:2022cwgv}.

\begin{figure}[tbhp]
\begin{center}
 \includegraphics[width=8cm]{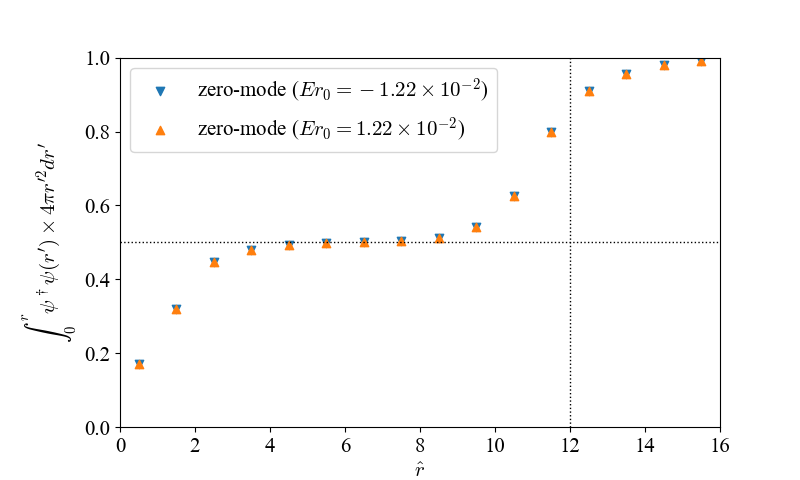}
 \includegraphics[width=8cm]{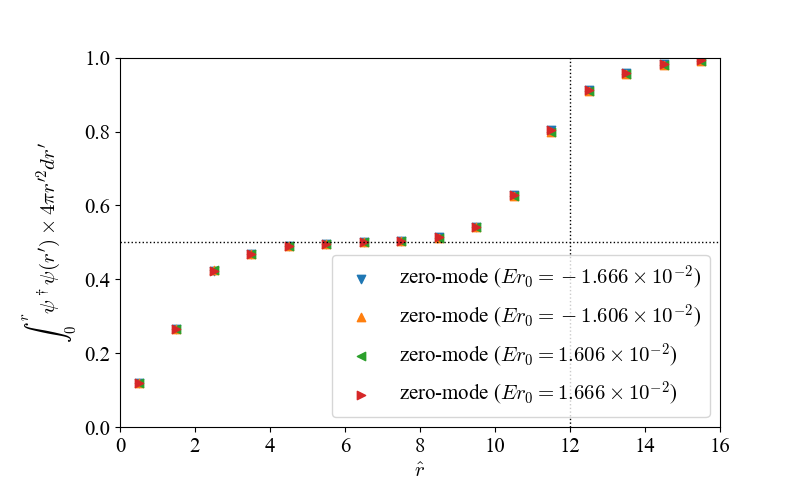}
\end{center}
\caption{Cumulative eigenfunction distribution of the nearest-zero modes inside $r$,
which gives an estimate for the electric charge captured by the monopole. 
The case with $n=1$ (left panel) and that with $n=-2$(right) are plotted.
At $4<r<9=3r_0/4$ we observe a stable plateau at 1/2.
 }
\label{fig:charge}
\end{figure}

Finally let us quantify the electric charge that the monopole gains.
In Fig.\ref{fig:charge}, we plot the cumulative distribution
\begin{align}
C_k(r) = \int_{|\bm{x}|<r} d^3x \phi_k(\bm{x})^\dagger \phi_k(\bm{x}),
\end{align}
of the $2|n|$ nearest-zero modes 
with magnetic charge  $n=1$ (left panel) and $n=-2$ (right).
They all show a stable plateau in the middle range
$4<r<9=3r_0/4$ at $C_k(r)\sim \frac{1}{2}$.
Therefore, we can conclude that under the half-filling condition,
the monopole gains $|n|/2$ electric charge 
capturing the half of the occupied $|n|$ zero-mode states of the electron.



\section{Effective theory revisited and simulation with non-integral magnetic charge}
\label{sec:EFT2}

\subsection{Reinterpretation of the effective action description}
The above analysis suggests an interesting reinterpretation 
of the original argument in the effective
Lagrangian of the Maxwell theory in Sec.~\ref{sec:EFT}.
When we integrate out the electron fields
in the presence of a single monopole,
the effective action contains a position-dependent
$\theta$ term:
\begin{align}
S_\text{top.}
&= \frac{1}{32\pi^2}\int d^4x \theta(r)F_{\mu\nu}\tilde{F}^{\mu\nu},\\
\theta(r)&= \pi \Theta(r-r_2)\Theta(r_0-r)=\left\{
\begin{array}{cc}\pi&\mbox{for $r_2<r<r_0$}\\ 0&\mbox{otherwise} \end{array}\right.,
\end{align}
where $\Theta(r)$ denotes the step function.
As is in the previous section,
$r_2$ is the size of the positive mass region around the
monopole, while $r_0$ is the location of the surface
when the topological insulator is spherical\footnote{
The shape of the outer surface is not important here.}.

Then the Maxwell equation is modified as
\begin{align}
\partial_\mu F^{\mu\nu} = - \frac{1}{8\pi^2}\partial_\mu [\theta(r)\tilde F^{\mu\nu}].
\end{align}
When we evaluate the volume integral inside the topological insulator $r<r_0$,
we have
\begin{align}
  q_e=\int_{r<r_0} d^3x \nabla \cdot \bm{E} &= -\frac{\theta}{4\pi^2}\int_{r<r_0} d^3x \nabla \theta(r) \cdot \bm{B}
  \nonumber\\&
  = -\frac{\theta}{4\pi^2}\int_{r<r_0} d^3x \pi\delta(r-r_1) \bm{e}_r\cdot \bm{B}
  = -\frac{1}{2}.
\end{align}
The result is consistent with Eq.~(\ref{eq:Gauss}) but
it does not require  $\nabla \cdot \bm{B}\neq 0$,
provided contribution from the Dirac string can be safely neglected.

The same argument applies to the case with a two-dimensional vortex with $\alpha=1/2$.
The effective action
after integrating out the fermion
can be treated as the Chern-Simons action with
a position dependent level $k(r)=\Theta(r-r_1)\Theta(r_0-r)$
and the Maxwell equation becomes
\begin{align}
\partial_\mu F^{\mu\nu} = - \frac{1}{8\pi^2}\partial_\mu [k(r)\epsilon^{\mu\nu\rho}A_\rho].
\end{align}
The electric charge around the vortex
\begin{align}
  q_e=\int_{r<r_0} d^2x \nabla \cdot \bm{E} &= -\frac{1}{2\pi}\int_{r<r_0} d^2x\partial_r k(r)A_\theta
  \nonumber\\&= -\frac{1}{2\pi}\int_{r<r_0} d^2x\delta(r-r_1)A_\theta
  = -\frac{r_1}{2\pi}\int_0^{2\pi} d\theta A_\theta = -\frac{1}{2},
\end{align}
is again consistent with Eq.~(\ref{eq:Gauss2D}) but
it does not require any singularity of the gauge field.

\subsection{Non-integral magnetic charge}

Knowing the fact that the singularity of the gauge field: $\nabla \cdot \bm{B}\neq 0$
is not necessarily required to gain the electric charge,
we would like to propose an interesting (thought) experiment
that may allow us to observe the Witten effect
without true monopoles.

Suppose we have a thin solenoid whose radius is comparable to
the crystal spacing of a topological insulator.
Inserting one end of the solenoid inside the topological insulator
while the other end is put outside, we can
mimic the monopole-anti-monopole system with a fine-tuned
electric current around the solenoid.
The solenoid represents the Dirac string.

\begin{figure*}
\includegraphics[width=0.49\textwidth, page=1]{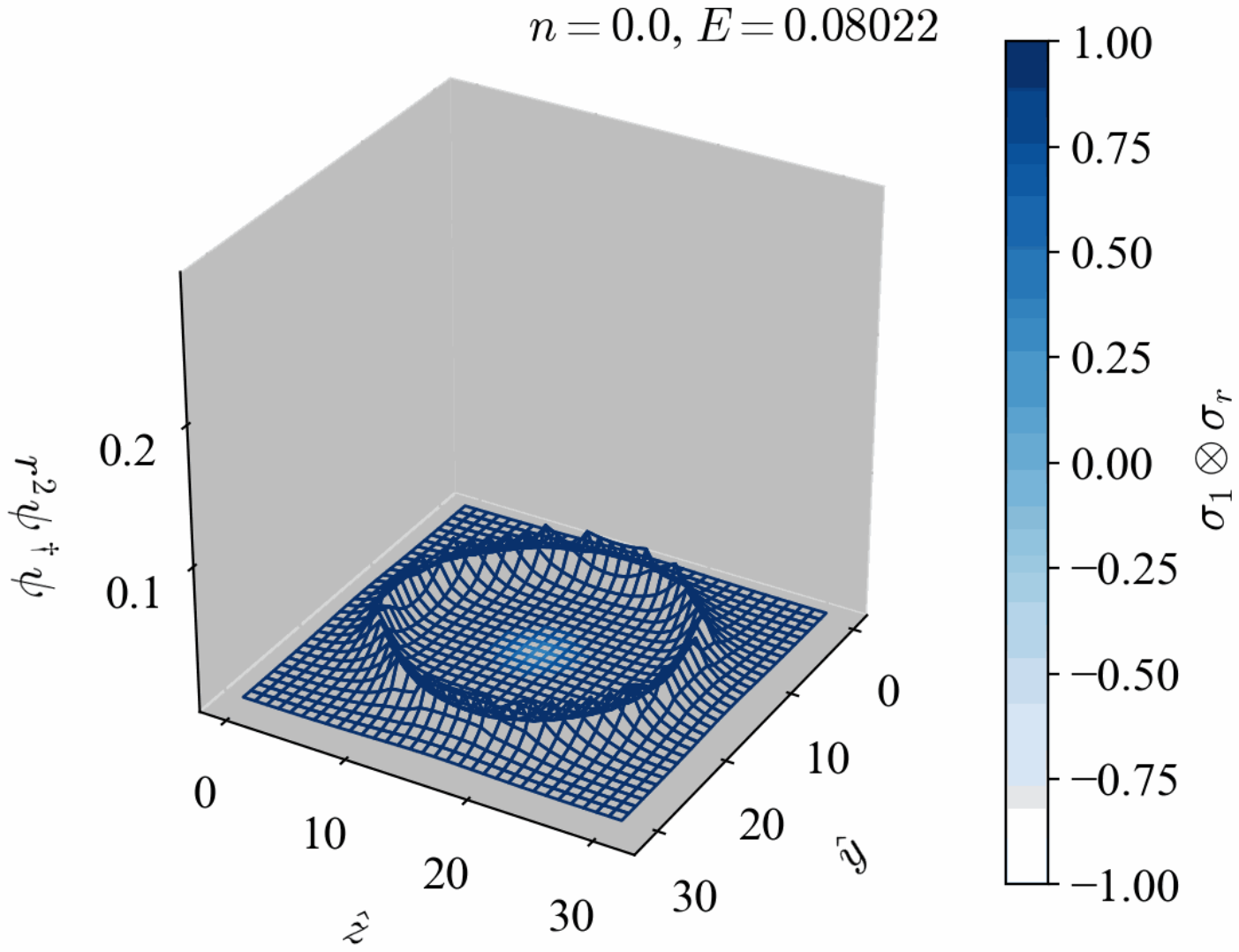}
\includegraphics[width=0.49\textwidth, page=2]{figures/L=32_r0=12_yzslice_Blues_selected.pdf}
\includegraphics[width=0.49\textwidth, page=3]{figures/L=32_r0=12_yzslice_Blues_selected.pdf}
\includegraphics[width=0.49\textwidth, page=4]{figures/L=32_r0=12_yzslice_Blues_selected.pdf}
\includegraphics[width=0.49\textwidth, page=5]{figures/L=32_r0=12_yzslice_Blues_selected.pdf}
\includegraphics[width=0.49\textwidth, page=6]{figures/L=32_r0=12_yzslice_Blues_selected.pdf}
  \caption{The amplitude (weighted by $r^2$) of the nearest-zero eigenfunction of $H_W$
at the slice $x=(L+1)/2$.   
By a continuous deformation of the monopole charge from $n=0$ to $n=1$,
    a half of the wave function goes into the vicinity of the monopole through the Dirac string.}
  \label{fig:varyingn}
\end{figure*}

This experiment can be simulated with our lattice setup
continuously varying the value of $n$ from 0 to 1.
The $n$ dependence of the amplitude weighted by $r^2$ 
of the positive nearest zero mode at the slice $x=(L+1)/2$
is presented in Fig.~\ref{fig:varyingn}.

At $n=0$ the energy eigenvalue is nonzero $E\sim 0.08$
and the amplitude is uniformly distributed around the
sphere of radius $r=r_0$.
Increasing $n$, a part of the wave function is gradually swallowed
into the topological insulator from bottom of the sphere.
At $n=0.5$, the amplitude along the Dirac string reaches maximum, 
and the energy eigenvalue is reduced to $E\sim 0.05$.
For larger $n>0.5$ the wave function is separated into two,
one half is attracted by the monopole, while
the other half stays at the original domain-wall. 
At $n=1$, the energy is reduced to almost zero.

In this process let us focus on a two-dimensional
slice at a fixed $z$ coordinate along the Dirac string.
The slice corresponds to the two-dimensional 
topological insulator with a vortex discussed in Sec.~\ref{sec:vortex}.
Note that the  flux of the vortex is equal to the monopole charge $\alpha=n$.
In Sec.~\ref{sec:vortex}, we have learned that the energy of the electron 
is minimized at $\alpha=1/2$ where the Aharanov--Bohm (AB) effect is maximized
and the wave function is localized at the vortex.
This explains why the amplitude of the electron at $n=0.5$
is attracted by the Dirac string.
At $\alpha=n=1$, the AB effect becomes zero and
the Dirac string cannot attract the electron field any longer,
while the binding force from the monopole is maximized
to capture the half of the electron field amplitude.

For $1<n<2$ a similar cycle starts on the second nearest zero mode.
The same continues for $n>3$ and the charge proportional to $|n|/2$
is, thus, pumped from the outer domain-wall to the monopole.

\begin{figure}[tbhp]
\begin{center}
 \includegraphics[width=12cm]{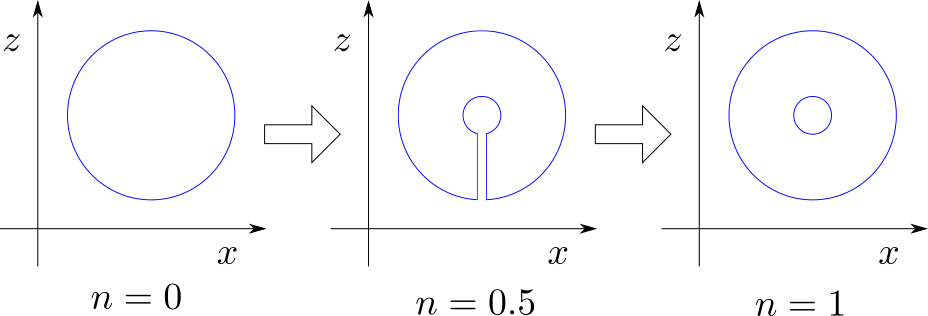}
\end{center}
\caption{
Schematic image of topology change of the domain-wall.
One spherical domain-wall at $n=0$ is extended
via the Dirac string into the location of the monopole at $n\sim 0.5$,
and is separated into two domain-walls at $n=1$. 
 }
\label{fig:topology-change}
\end{figure}

It is also interesting to trace the topology change
of the domain-wall shown in Fig.~\ref{fig:topology-change}.
Increasing $n$ from zero to $1/2$,
the mass shift along the Dirac string becomes large,
making a bridge of normal insulators connecting the
$r>r_0$ region and the location of the monopole.
By the time the monopole charge  reaches $n=1$,
 the Dirac string neighbor
becomes back to the topological insulator again,
and the bridge is disconnected to end up with
two spherical domain-walls: one at the original $r=r_0$
and the other at the vicinity of the monopole.

\section{Summary and discussion}
\label{sec:summary}

In this work, we have investigated the microscopic mechanism
how a vortex/monopole in topological insulators 
becomes a dyon gaining a half-integral electric charge.
In our analysis we have added the Wilson term to the Dirac equation,
compared to which the sign of the electron mass has a physical meaning.
We have also smeared the singularity of the gauge field 
in a finite region of radius $r_1$,
which makes our  computations always UV finite.
In this set up, we have obtained the (near-)zero mode solutions
both analytically in continuum
and numerically on a lattice.

The key finding in this work is dynamical creation
of a domain-wall around the monopole/vortex.
We have confirmed that 
the local mass shift from the Wilson term
is of size $1/r_1$ that is positive.
This mass shift is strong enough to
make a small island of a normal insulator around
the monopole/vortex.
The edge-localized modes on the domain-wall 
can be identified as the origin of the electric charge
captured by the monopole/vortex.
In the $r_1=0$ limit with a quantized magnetic charge $q_m=1/2 \; \mbox{mod}\; \mathbb{Z}/2$
or flux $\alpha=1\; \mbox{mod}\; \mathbb{Z}$, 
only the chiral modes survive with the energy zero.

The fact that the created domain-wall is not 
a point-like singularity but a codimension-one 
sphere having the same dimensions that 
the outer surface of the target topological insulator has,
makes the topological meaning of the zero modes clearer.
In mathematics, the two manifolds of the same dimension are
cobordant if their disjoint union is the boundary
of a compact manifold one dimension higher.
In our physics target, it is clear that 
the dynamically created small domain-wall at $r=r_1$ around the vortex
or the one at $r=r_2\sim 1/M_\text{PV}$ around the monopole,
and the surface at $r=r_0$ 
of the topological insulator are cobordant since 
they are the boundaries of the same topological insulator 
having a negative mass.
We have also identified the number of the chiral zero modes
as the Atiyah-Singer or mod-two Atiyah-Singer index, 
which is a cobordism invariant:
they must appear in pairs on the domain-wall $r=r_1$ or $r_2$
and on the one at $r=r_0$.

For each set of the two zero modes paired by the 
cobordism, we have computed their mixing amplitude 
both analytically and numerically.
Due to the chiral symmetry of the total system,
at whatever large distance $r_0$, 
the mixing is always maximum: 50\% vs. 50\%,
which explains why the electric charge expectation value
around the monopole/vortex is 1/2 in the half-filling state,
where only one of the two split near zero modes is occupied.

From our microscopic analysis we have suggested 
an interesting reinterpretation of the effective theory approach.
After integrating out the massive electron fields,
the system contains a defect not in the gauge field but
in the $\theta$ parameter or the level $k$ of the Chern-Simons action.
The modified Maxwell equation still gives the
same half-integral electric charge 
but does not require any singularity of the gauge field,
since the monopole/vortex 
is located in the trivial $\theta=0$ or $k=0$ region.

This implies that the Witten effect can be observed 
in our laboratory even  without true monopoles.
We have simulated a thin solenoid, which corresponds to 
the Dirac string, with one end inserted into 
a topological insulator and the other end outside.
Increasing the magnetic flux of the solenoid, 
we have observed that
a bridge of normal insulators along the solenoid  appears 
between the surface and the monopole.
Then a pair of the near-zero edge modes on the surface
gradually crosses the bridge towards the monopole
decreasing its energy to zero.
By the time the magnetic flux reaches that of 
a unit magnetic monopole, a half of the wave function
arrives at the monopole, and the bridge disappears.
In this process, a single domain-wall sphere 
is separated into two, to populate a pair of chiral zero modes.

It is interesting to investigate how stable
our results are against perturbation of the Hamiltonian
due to impurities or external interference.
While in the original discussion in Ref.~\cite{Witten:1979ey} 
the half integral charge depends only on 
the $T$-symmetric value of $\theta=\pi/2$, our analysis 
has relied on an additional chiral symmetry with $\bar\gamma$
to show topological stability of the zero modes.
It is a good challenge to examine if breaking of the chiral symmetry
keeping the $T$ symmetry makes the results unchanged.

We thank Mikio Furuta, Masahiro Hotta, Takuto Kawakami, Tsunehide Kuroki, Okuto Morikawa,  
Tetsuya Onogi, Shigeki Sugimoto, Kazuki Yamamoto, Masahito Yamazaki, and Ryo Yokokura for useful discussions.
The work of SA was supported by JST SPRING, Grant Number JPMJSP2138.
The work of HF and NK was supported by JSPS KAKENHI Grant Number JP22H01219.
The work of MK was supported by JSPS KAKENHI Grant Number JP21H05236, JP21H05232, 
JP20H01840, and JP20H00127, and by JST CREST Grant Number JPMJCR20T3, Japan.

\bibliography{ref}

\end{document}